\documentclass[twocolumn]{aastex631}  
\usepackage{enumitem}
\usepackage{xcolor}
\usepackage{tablefootnote}
\usepackage{bm}
\usepackage{lineno}

\hypersetup{backref=true,  pagebackref=true,  hyperindex=true, colorlinks=blue,	breaklinks=true,  urlcolor=violet,  linkcolor= red, bookmarks=true, 	bookmarksopen=true,  filecolor=cyan, citecolor=blue, linkbordercolor=blue}

\newcommand\wfirst{\textit{Roman}}
\shorttitle{Parallax in astrometric microlensing}
\shortauthors{Sajadian, et. al., }

\begin{document}
\title{Discerning Parallax Amplitude in Astrometric Microlensing}
\author[0000-0002-0167-3595]{Sedighe~Sajadian$^{a}$}
\author{Arya Mahmoudzadeh}
\author{Setareh Moein}
\affiliation{Department of Physics, Isfahan University of Technology, Isfahan 84156-83111, Iran}
\footnote{$^{a}$\textcolor{blue}{$\rm{e}$-$\rm{mail:s.sajadian@iut.ac.ir}$}}

\begin{abstract}
Gravitational microlensing is a powerful method for discovering Isolated Stellar-Mass Black Holes(ISMBHs). These objects make long-duration microlensing events. To characterize these lensing objects by fully resolving the microlensing degeneracy, measurements of parallax and astrometric deflections are necessary. Microlensing events due to ISMBHs have considerable astrometric deflections, but small parallax amplitudes as $\pi_{\rm E} \propto 1/\sqrt{M_{\rm l}}$, where $M_{\rm l}$ is the lens mass. We numerically investigate the possibility of inferring parallax amplitude from astrometric deflection in microlensing events due to ISMBHs. The parallax amplitude in astrometric deflections is proportional to the relative parallax $\pi_{\rm{rel}}$, which means (i) does not strongly depend on $M_{\rm l}$, and (ii) increases in microlensing observations toward the Magellanic Clouds(MCs). We assume these events are potentially detected in upcoming microlensing surveys-(1): the \wfirst\ observations of the Galactic bulge (GB), and (2): the LSST observations of the Large MC(LMC)-, and the Extremely Large Telescope (ELT) follows up them with one data point every ten days. We evaluate the probability of inferring parallax amplitude from these observations by calculating the Fisher/Covariance matrices. For GB, the efficiencies for discerning parallax amplitudes with a relative error $<4\%$ through astrometric, and photometric observations are $3.8\%$, and $29.1\%$, respectively. For observations toward the LMC, these efficiencies are $41.1\%$, and $23.0\%$, respectively. Measuring parallax amplitude through astrometric deflections is plausible in the GB events with the lens distance $\lesssim 2.7$kpc, and in the LMC halo-lensing. The ELT telescope by monitoring long-duration microlensing events can detect astrometric deflections, and their parallax-induced deviations.
\end{abstract}


\section{Introduction}
A short time after the famous paper by \citet{1986Paczynski} concerning the detection of Massive Compact Halo Objects (MACHOs) in the Galactic halo through continuous observations toward the Magellanic Clouds (MCs), three microlensing groups, i.e., the Exp\'erience de Recherche d' Objets Sombres (EROS, \citet{1993EROS}), MACHO \citep{1994Machogroup}, and Optical Gravitational Lensing Experiment (OGLE, \citet{1994oglegroup}), searched for ongoing microlensing events. The first generation of microlensing surveys observed MCs for $10$ years to find the source stars that were being lensed by massive objects inside the Galactic halo. The most important result of these observations was the determination of an upper limit on the contribution of MACHO in the Galactic halo \citep{2000Erosres,1998Alcock, 2000MachoLMC}. Since MACHOs have only gravitational interactions, so studying and characterizing these objects are possible through gravitational microlensing observations.

The second generation of survey microlensing groups changed observational directions from MCs to the Galactic Bulge (GB) and spiral arms to discover extra solar planets inside the Galactic disk, and additionally probe the Galactic structure \citep{1994GalacticMpacz,1996planetCol,2000OGLEbulge,1999AAErosII, 2000MachoBulge,Marc2017}. In planetary microlensing events, an extra solar planet (exoplanet) is usually orbiting the lens object and makes a deviation in the magnification curve \citep[see, e.g., ][]{1991Maoplanet,1992ApJgould}. So far, more than $204$ extra solar planets have been confirmed in microlensing observations\footnote{\url{https://exoplanetarchive.ipac.caltech.edu/}}. This population of exoplanets discovered by gravitational microlensing observations has the special properties listed here. (a) These exoplanets lie mostly beyond the snowline of their parent stars. (b) In microlensing events toward the GB, planetary lensing systems are typically found at the radial distances greater than $1000$ parsec from the observer. (c) These exoplanets are orbiting the host stars which are usually very dim or dark \citep[see, e.g., ][]{2012Gaudireview}. Gravitational microlensing is therefore a complementary method to other methods for discovering extrasolar planets. In other methods (e.g., transits, radial velocimetry, astrometry), planetary systems must be close to the observer and their host stars must be bright.

Applications of survey microlensing observations are not limited to the mentioned cases. For example, a new population of free-floating planets (FFPs) in the Galactic disk was discovered through densely observing the GB \citep{2023AJSumi,2017NaturMroz}. Discovering Isolated Stellar-Mass Black Holes (ISMBHs) inside the Galactic disk is another outstanding application of gravitational microlensing  \citep{2022ApJSahu,2022ApJLam,2002MNRASMao,2002Bennett,2002ApJAgol,2016ApJLu}. Such isolated objects most likely do not emit $X$-ray emissions. Hence, they are only discernible through probing long-duration and achromatic amplifications in the light of background source stars.

Despite such vast and unique applications of gravitational microlensing, there is a problem while interpreting microlensing observations, which is degeneracy \citep[see, e.g., ][]{2023MNRASsajadian}. However, microlensing degeneracies can be resolved through measuring both (a) the parallax amplitude in magnification factor \citep{Gould1994}, and (b) finite-source effect \citep{1994Witt}. Instead of resolving finite-source effect, measuring either lensing-induced astrometric deflection in source trajectory \citep[see, e.g., ][]{1995Hog,1995Miyamoto,1995Walker,1996MialdaAstrometric,2000ApJDominikSahu}, or the images' distance at the time of the closest approach \citep{2019ApJDong,2020ApJZang}, or resolving the source and lens, and measuring proper motions with adaptive optics observations \citep{2018AJBhattacharya,2022AJTerry} will help to resolve the microlensing degeneracy. The parallax effect refers to the observer's motion around the Sun. This effect makes some periodic perturbations in microlensing light curves \citep{Gould1994}.

In long-duration microlensing events due to dark and massive lens objects (e.g., ISMBHs), this method for resolving microlensing degeneracy is somewhat challenging, because the normalized parallax amplitude decreases like $\pi_{\rm E}\propto 1/\sqrt{M_{\rm l}}$, where $M_{\rm l}$ is the lens mass \citep[see, e.g., ][]{2020Karolinski}. Albeit, the scale of the astrometric deflection in the source trajectory (i.e., $\theta_{\rm E}$ which is the angular Einstein radius) in these events is considerable, because $\theta_{\rm E}\propto \sqrt{M_{\rm l}}$. In this work, we discuss on the possibility of discerning parallax amplitudes in astrometric deflections of source trajectories. We show that parallax-induced perturbations in the astrometric deflection are directly proportional to the relative parallax $\pi_{\rm{rel}}$, and do not depend on the lens mass strongly. Hence, measuring parallax amplitude is possible in the astrometric deflections of source trajectories in microlensing events toward MCs. However, microlensing events due to more massive lens objects are more suitable, because they have considerable astrometric deflections in source trajectories.

The paper is organized as follows. In the first subsection of Section \ref{astro}, we first explain the formalism of astrometric microlensing events by including the parallax effect. Then, in subsections \ref{simul1} and \ref{simul2}, by simulating astrometric microlensing events due to ISMBHs by considering the parallax effect toward the GB and LMC (respectively) we statistically evaluate parallax-induced deviations. In Section \ref{mcmc}, we do realistic Monte Carlo simulations based on upcoming survey microlensing observations by the \wfirst \citep{2015Spergel} and LSST \citep{lsstbook} telescopes. We also consider potential follow-up observations with the Extremely Large Telescope (ELT) telescope by taking one data point every ten days \citep{2011PASPELT}. In these simulations, we numerically calculate the Fisher and Covariance matrices to evaluate the probability of discerning parallax amplitudes through astrometric and photometric observations. In Section \ref{result}, we explain the results and conclude.

\section{Parallax effect in astrometric mirolensing} \label{astro}
In this section, we aim to answer this question: "In what kind of microlensing events are the parallax amplitudes in astrometric deflections more realizable than those in the magnification curves?" For this aim, in Subsection \ref{forma} we first review our formalism for generating astrometric microlensing events by considering the parallax effect. Then, in Subsections \ref{simul1} and \ref{simul2} we simulate these events, and statistically evaluate parallax-induced perturbations. 

\subsection{Formalism}\label{forma}
In a gravitational microlensing event, the light of a background star is temporarily magnified due to passing through the gravitational potential of a collinear and foreground massive object \citep{Einstein1936}. In this phenomenon from a background star two images are formed whose angular positions and magnification factors are (respectively):  
\begin{eqnarray}
\boldsymbol{\theta}_{\pm}= \frac{u\pm \sqrt{u^{2}+4}}{2} \theta_{\rm E}\hat{u}, ~~A_{\pm}= \frac{1}{2}\Big(\frac{u^{2}+2}{u\sqrt{u^{2}+4}} \pm 1\Big),
\end{eqnarray}
where, $u$ is the angular distance of the source star from the lens object normalized to the angular Einstein radius (i.e., $\theta_{\rm E}$ the angular radius of the images' ring when the lens, source star and observer are completely aligned). $\hat{u}$ is a unit vector represents the direction of $u$ projected on the sky plane.

\noindent Hence, the locations of two images and the source star are over a straight line in the sky plane. The angular distance of these images ($\simeq 2\theta_{\rm E}$) is too small to be resolved, because $\theta_{\rm E}$ is on milliarcsecond scales. For instance, for a common lens object (e.g., an M-dwarf star) in the Galactic disk ($D_{\rm l}\simeq 4$ kpc, where $D_{\rm l}$ is the lens distance from the observer) while the source star is inside the GB (i.e., the source distance from the observer is $D_{\rm s}\simeq 8$ kpc), it is given by
\begin{eqnarray}
\theta_{\rm E}=\sqrt{\kappa~ \pi_{\rm{rel}}~M_{\rm l}}=0.78\rm{(mas)} ~\sqrt{\frac{M_{\rm l}}{0.3 M_{\odot}}}\sqrt{ \frac{\pi_{\rm{rel}}}{0.25~\rm{mas}}}, 
\label{tetE}
\end{eqnarray}
here, $\pi_{\rm{rel}}=\rm{au}\big(1/D_{\rm l}-1/D_{\rm s}\big)$ is the so-called relative parallax, au is the astronomical unit, and $\kappa=8.14~\rm{mas}~M_{\odot}^{-1}$ has a constant value. Therefore, we receive the total light of images which is magnified, the so-called gravitational microlensing event. The magnification factor due to both images is given by:  
\begin{eqnarray}
A=\frac{u^{2}+2}{u\sqrt{u^{2}+4}}.
\label{magni}
\end{eqnarray}

In a microlensing event, in addition to the light magnification of a background star, the brightness center of these images does not coincide to the source's center. The astrometric deflection in the source position is given by \citep{1995Hog,1995Miyamoto,1995Walker,1996MialdaAstrometric}:   
\begin{eqnarray}
\delta \boldsymbol{\theta}=\frac{\boldsymbol{\theta}_{+}A_{+} + \boldsymbol{\theta}_{-}A_{-}}{A_{+}+A_{-}}- \boldsymbol{u}~\theta_{\rm E}= \frac{\boldsymbol{u}~\theta_{\rm E}}{u^{2}+2}, 
\label{shift}
\end{eqnarray}
which is a function of time. This shift is a dimensional parameter, and could be used for measuring the angular Einstein radius. 

By ignoring the motion of the observer around the Sun (i.e., we first assume the observer is the Sun), the vector of the relative lens-source position in the heliocentric frame, $\boldsymbol{u}_{\odot}$, is expressed as a function of time $t$: 
\begin{eqnarray}
\boldsymbol{u}_{\odot}= 
 \left(
 \begin{array}{cc}
(t-t_{0})/t_{\rm E}\\
u_0  
\end{array}     \right),
\end{eqnarray} 
where, $u_{0}$ is the lens impact parameter (the closest lens-source distance), $t_{0}$ is the time of the closest approach, and $t_{\rm E}=\theta_{\rm E}/\mu_{\rm{rel}, \odot}$ is the so-called Einstein crossing time, the time of crossing the angular Einstein radius. Here, $\mu_{\rm{rel}, \odot}$ is the angular lens-source relative velocity as measured in the heliocentric frame.  

We now add the Earth motion around the Sun (by assuming the observer is on the Earth), the so-called parallax effect. In gravitational microlensing events, the parallax effect causes the angular lens-source relative velocity alters as 
\begin{eqnarray}
\boldsymbol{\mu}_{\rm{rel}}=\boldsymbol{\mu}_{\rm{rel},~\odot} + \frac{\pi_{\rm{rel}}}{\rm{au}} \boldsymbol{v}_{\rm o, \perp}(t),
\label{mul}
\end{eqnarray}        
where $\boldsymbol{v}_{\rm o, \perp}$ is the vector of the Earth velocity with respect to the Sun projected on the sky plane (normal to the line of sight). Therefore, the parallax effect changes the vector of the relative lens-source distance as measured from the Earth as 
\begin{eqnarray}
\boldsymbol{u}= \boldsymbol{u}_{\odot} + \frac{\pi_{\rm E}}{\rm{au}} \int_{t_{0}}^{t} dt~ \boldsymbol{v}_{\rm o, \perp}(t)=\boldsymbol{u}_{\odot}+\pi_{\rm E} \boldsymbol{\Delta}_{\rm o,~n}(t),
\label{eqqq}
\end{eqnarray}
here, $\pi_{\rm E}=\pi_{\rm{rel}}/\theta_{\rm E}$ is the relative lens-source parallax normalized to the angular Einstein radius. $\boldsymbol{\Delta}_{\rm o,~n}$ is an extra displacement in the source position with respect to the lens due to the Earth motion, normalized to the astronomical unit. 

\noindent We note that in real observations the Earth motion around the Sun has been completely known, but (i) $\pi_{\rm E}$, and (ii) the angle between the Earth velocity projected on the sky plane and the source trajectory are unknown. For that reason parallax effect includes two new variables in the lensing formalism, i.e., $\pi_{\rm E}$, and $\xi$. Since, the relative source-lens trajectory projected on the sky plane has a fixed direction (i.e., the Sun, and other stars move on straight lines during lensing time scales), so we define $\xi$ as the angle between $\boldsymbol{v}_{\rm o, \perp}$ and $\boldsymbol{\mu}_{\rm{rel},~\odot}$ at the time of closest approach $t_{0}$.
\begin{figure*}
\centering
\includegraphics[width=0.32\textwidth]{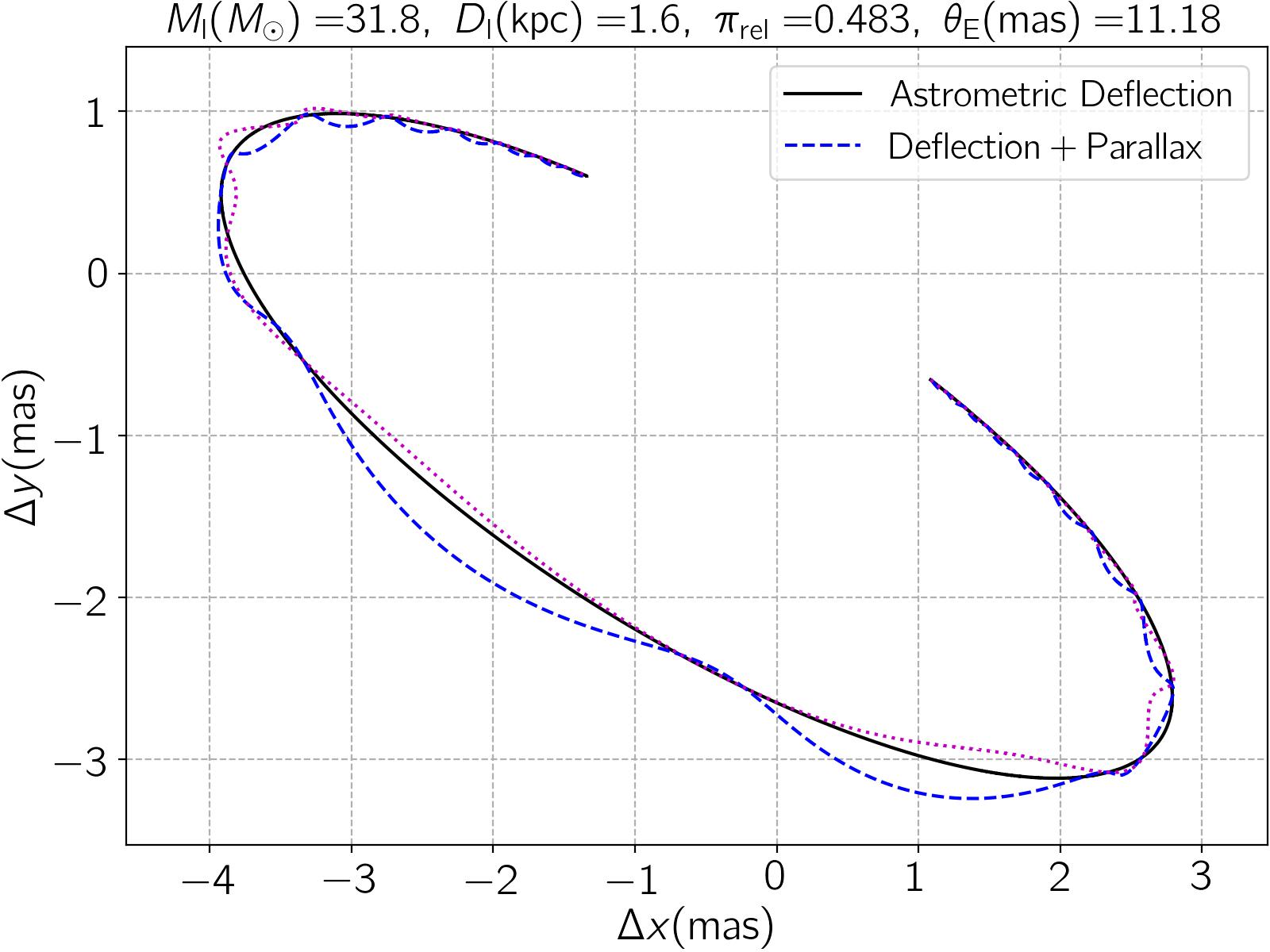}
\includegraphics[width=0.32\textwidth]{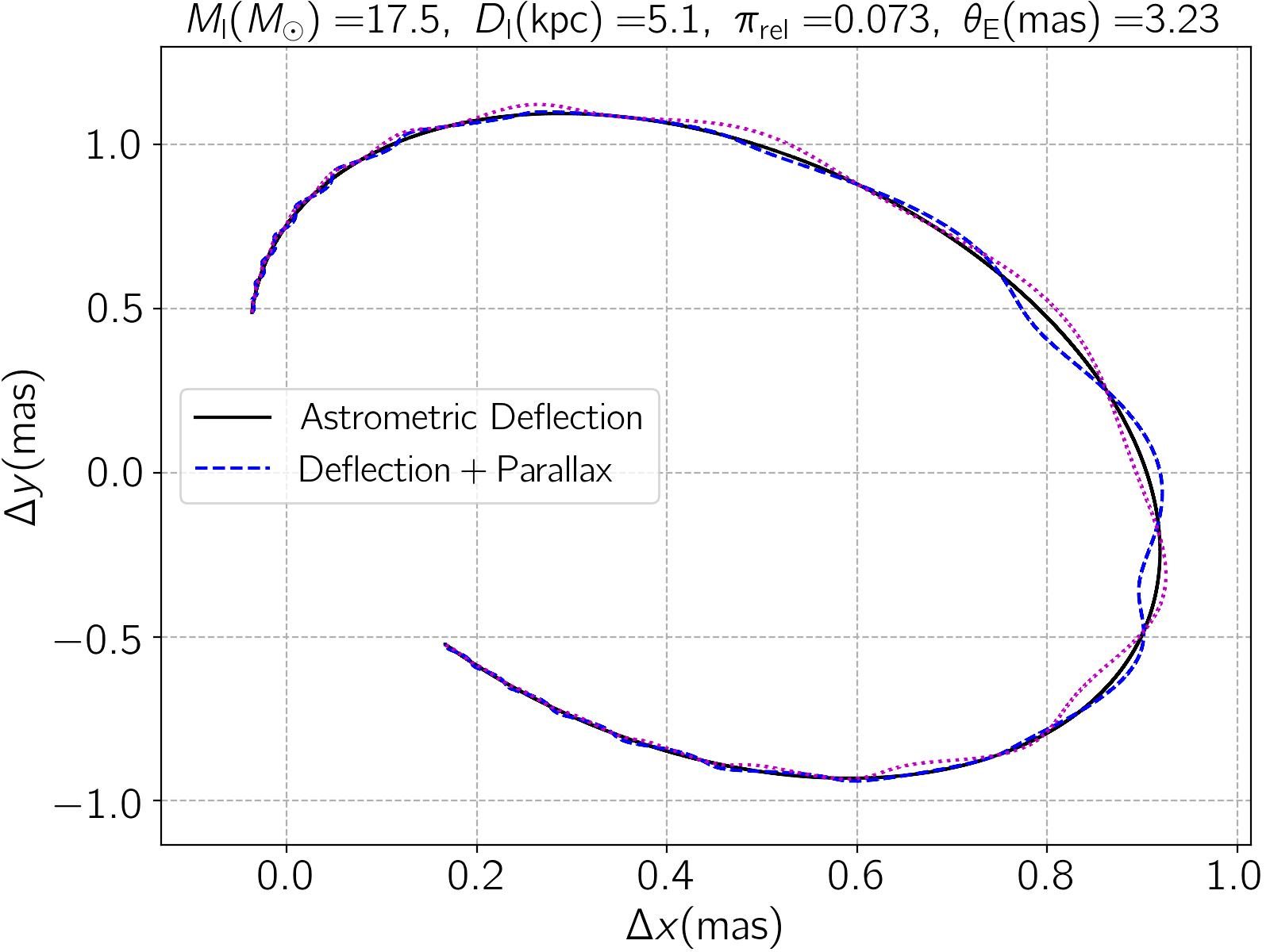}
\includegraphics[width=0.32\textwidth]{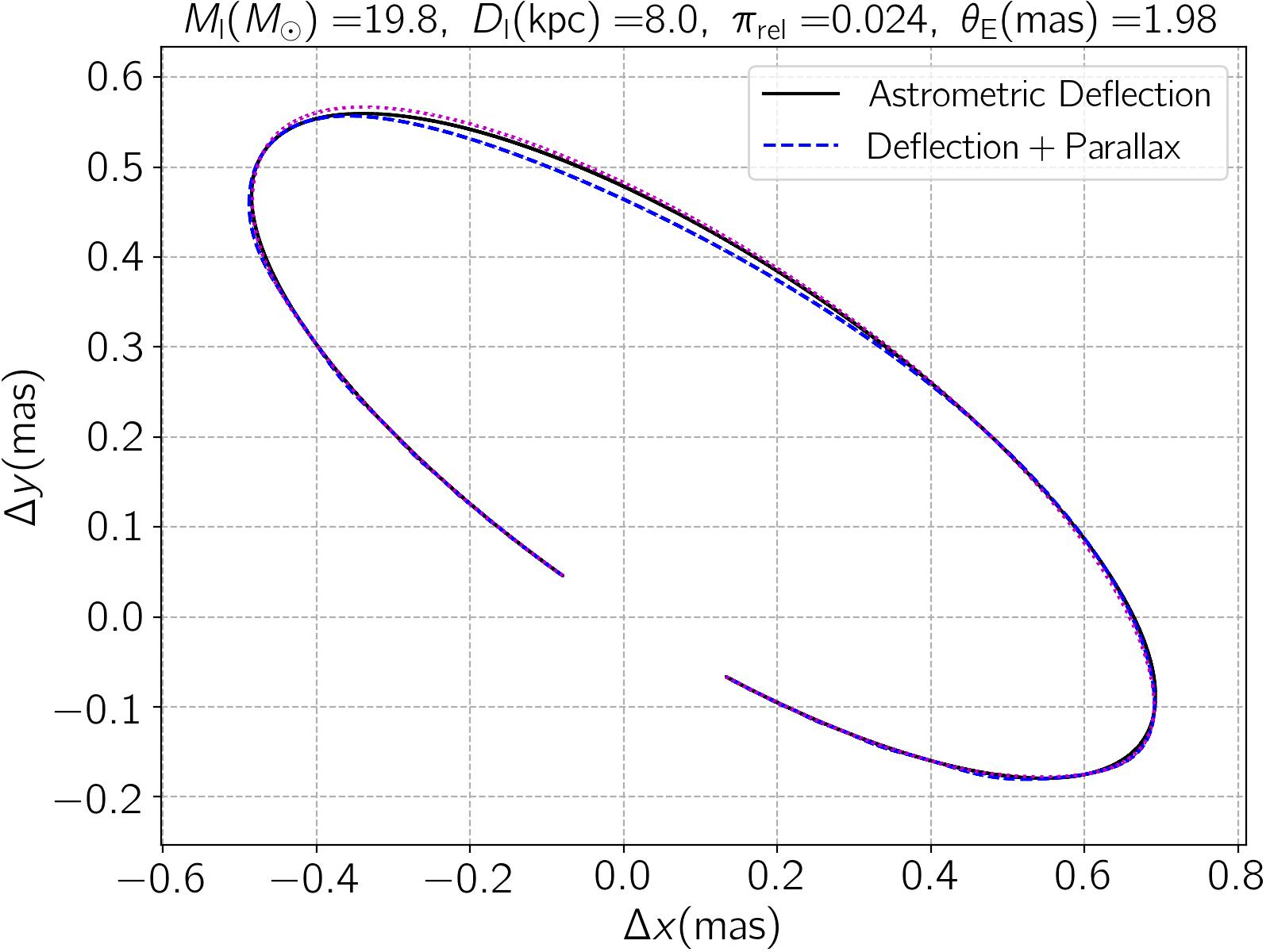}
\includegraphics[width=0.32\textwidth]{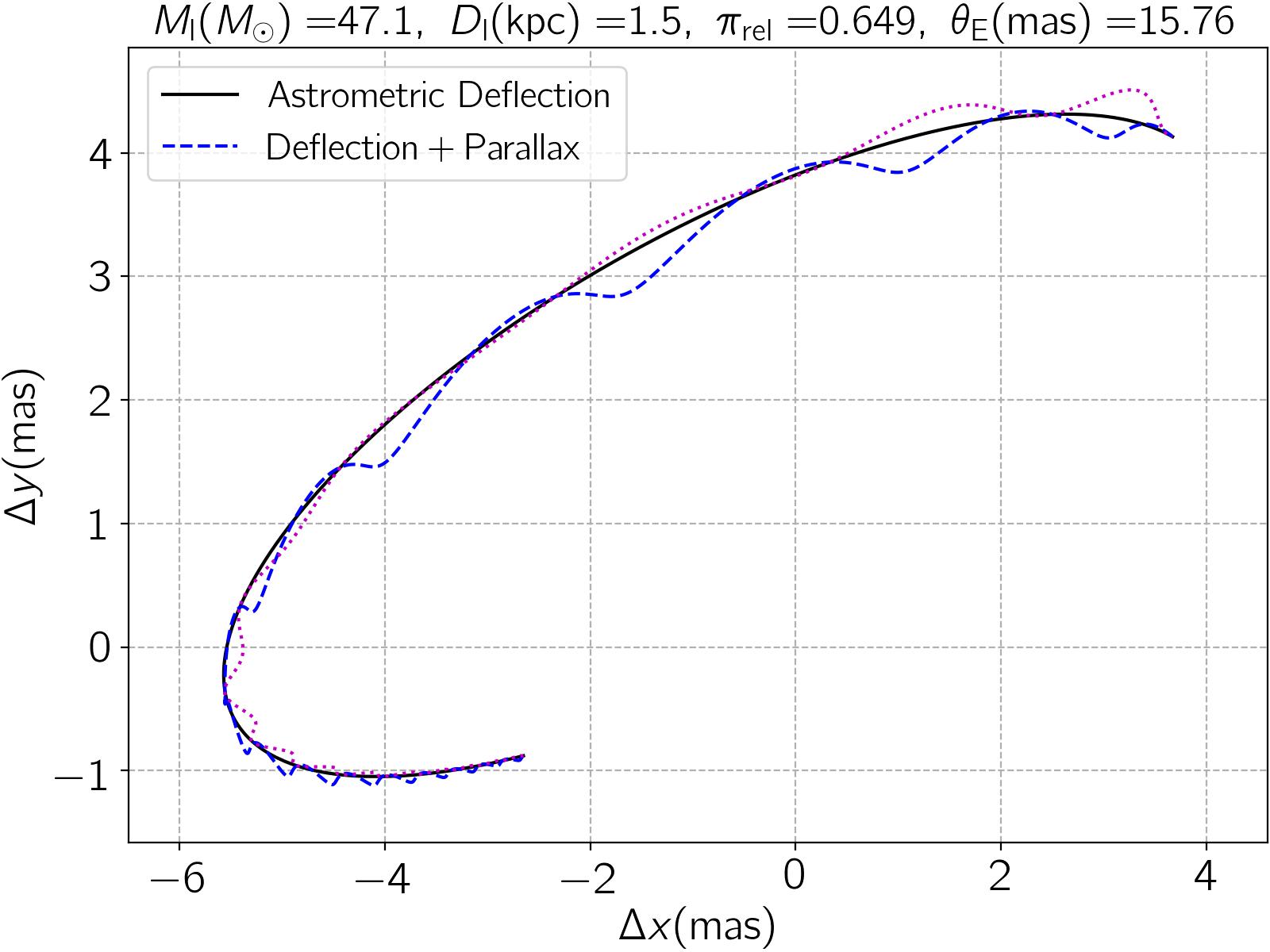}
\includegraphics[width=0.32\textwidth]{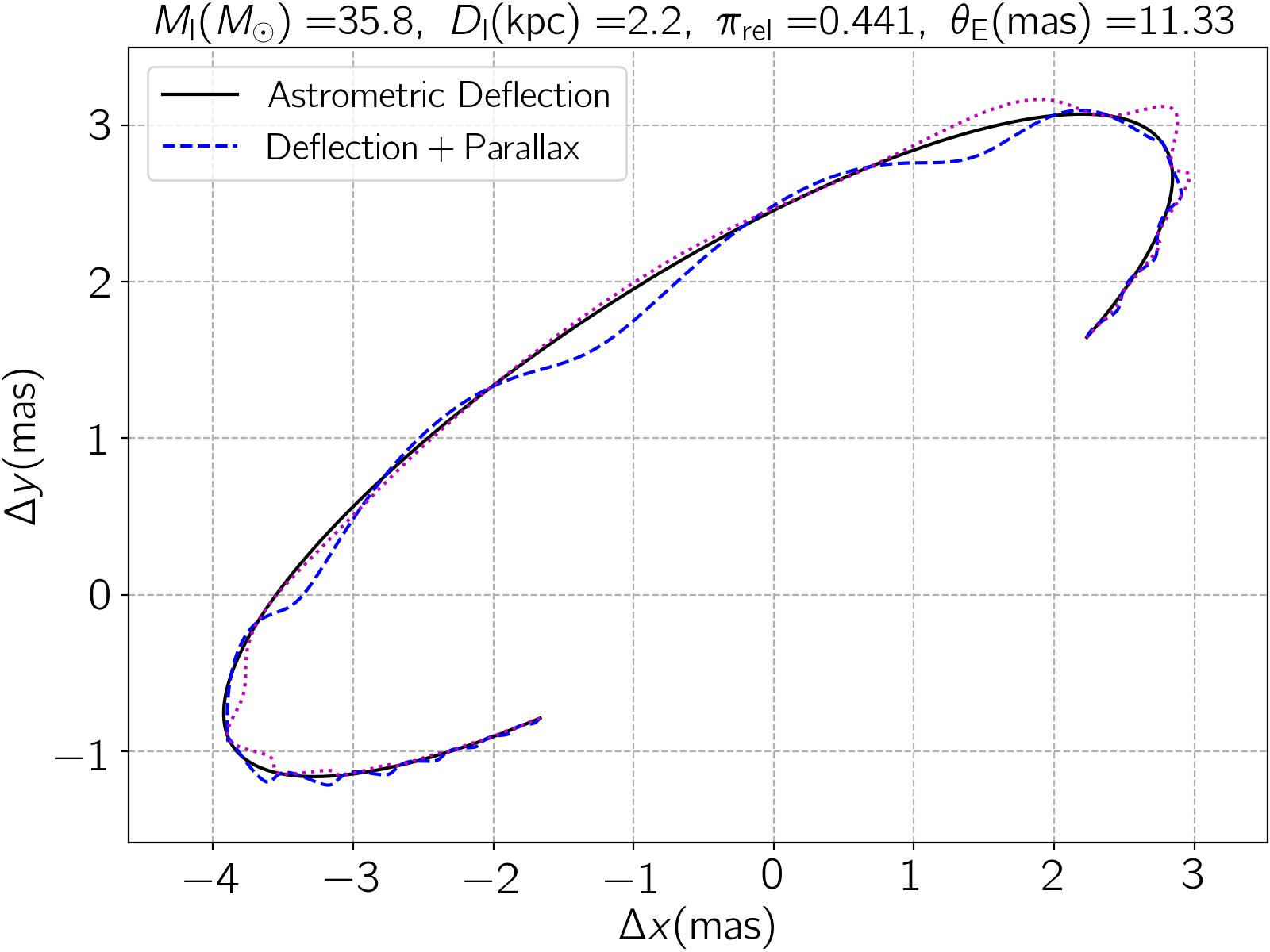}
\includegraphics[width=0.32\textwidth]{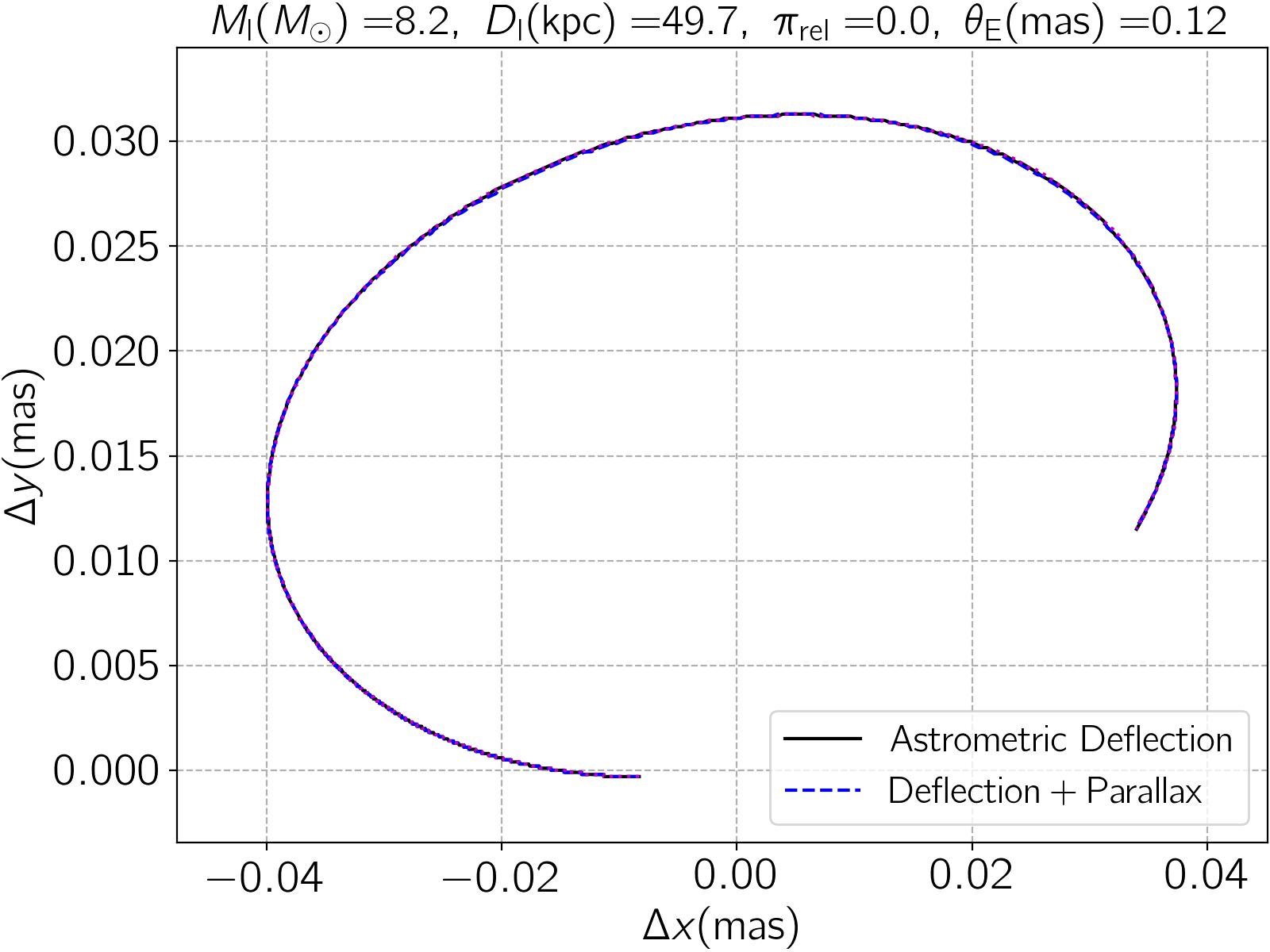}
\caption{Examples of astrometric deflections in source trajectories in long-duration microlensing events toward the GB (three top panels), and LMC (three bottom panels) due to ISMBHs, by ignoring (solid black curves) and including (dashed blue curves) parallax effect. The dotted magenta curves are the first sentence in the astrometric deflection given by Equation \ref{twoterm}, i.e., $\delta \boldsymbol{\theta}_{0}$. Their parameters were mentioned at the top of curves.}\label{Fig_shift}
\end{figure*}

The second term in Equation \ref{eqqq} changes periodically, and causes a periodic perturbation in the magnification curve and astrometric deflection in the source trajectory. The amplitude of this perturbation in the magnification factor is $\pi_{\rm E}$ because the magnification factor $A$ depends only on $u$.
  
\noindent The astrometric deflection in the source trajectory by considering the parallax effect is
\begin{eqnarray}
\delta \boldsymbol{\theta}=\delta \boldsymbol{\theta}_{0}+\delta \boldsymbol{\theta}_{1}= \frac{\boldsymbol{u}_{\odot}~\theta_{\rm E}}{u^{2}+2} + \frac{\boldsymbol{\Delta}_{\rm o, n}~\pi_{\rm{rel}}}{u^{2}+2}, 
\label{twoterm}
\end{eqnarray}
where, $u^{2}=u_{\odot}^{2}+ \pi_{\rm E}^{2}\Delta_{\rm o, n}^{2}+2~\pi_{\rm E}~ \boldsymbol{u}_{\odot}.\boldsymbol{\Delta}_{\rm o, n}$. The first sentence makes an ellipse on the lens plane (with small perturbations), and the second term makes a periodic perturbation over that ellipse. The amplitude of these periodic perturbations is proportional to $\pi_{\rm{rel}}$. However, the larger $\theta_{\rm E}$, the higher detectability of these perturbations. Because (i) the amplitude of the astrometric deflection is proportional to $\theta_{\rm E}$, and (ii) the higher $\theta_{\rm E}$ makes lower $\pi_{\rm E}$, and as a result, the second term in Equation \ref{twoterm} increases. According to Equation \ref{tetE}, long-duration microlensing events due to ISMBHs close to the observer (in comparison with the source distance) are the most suitable ones to realize the parallax effect in their astrometric deflections instead of magnification curves.

Briefly, the parallax effect changes both the magnification factor and the astrometric deflection in the source position (as given by Eq.  \ref{magni}, and \ref{shift}) in different ways, as:
\begin{itemize}[leftmargin=2.0mm]
\item These two observing features have different time scales, as they tend to zero by $u^{-4}$, and $u^{-1}$, respectively \citep[see, e.g., ][]{2000ApJDominikSahu,2014MNRASorbital}. By getting away the source star from the gravitational potential of the lens object, the astrometric deflection in the source position tends to zero very slowly. The slow evolution of the astrometric shift is beneficial to realize annual parallax amplitude.  
	
\item The amplitudes of parallax-induced perturbations in the magnification factor and the astrometric deflection are proportional to $\pi_{\rm E}$, and $\pi_{\rm{rel}}$, respectively. Considering the fact that for realizing the astrometric deflection itself, $\theta_{\rm E}$ should be large, in long-duration microlensing events due to ISMBHs very close to the observer the parallax amplitude could be discerned in astrometric deflections instead of magnification factors. 

\item When $u=u_{0}$, the magnification factor is maximum, whereas for $u=\sqrt{2}$ the astrometric deviation is maximum. Hence, discerning the parallax-induced perturbation in the magnification factor, and astrometric deflection can be done with a higher probabilities when $u\ll 1$ and $u\simeq 1$, respectively. This point can be found in Figure \ref{NDtot}. 
\end{itemize}

So searching the parallax effect in the astrometric deflection instead of the magnification factor would be possible in long-duration microlensing events due to massive and close lens objects (e.g., stellar-mass or intermediate-mass black holes). These points will be verified numerically in the next subsections by performing simulations of microlensing events toward the GB and LMC. 

\begin{figure*}
\centering
\includegraphics[width=0.32\textwidth]{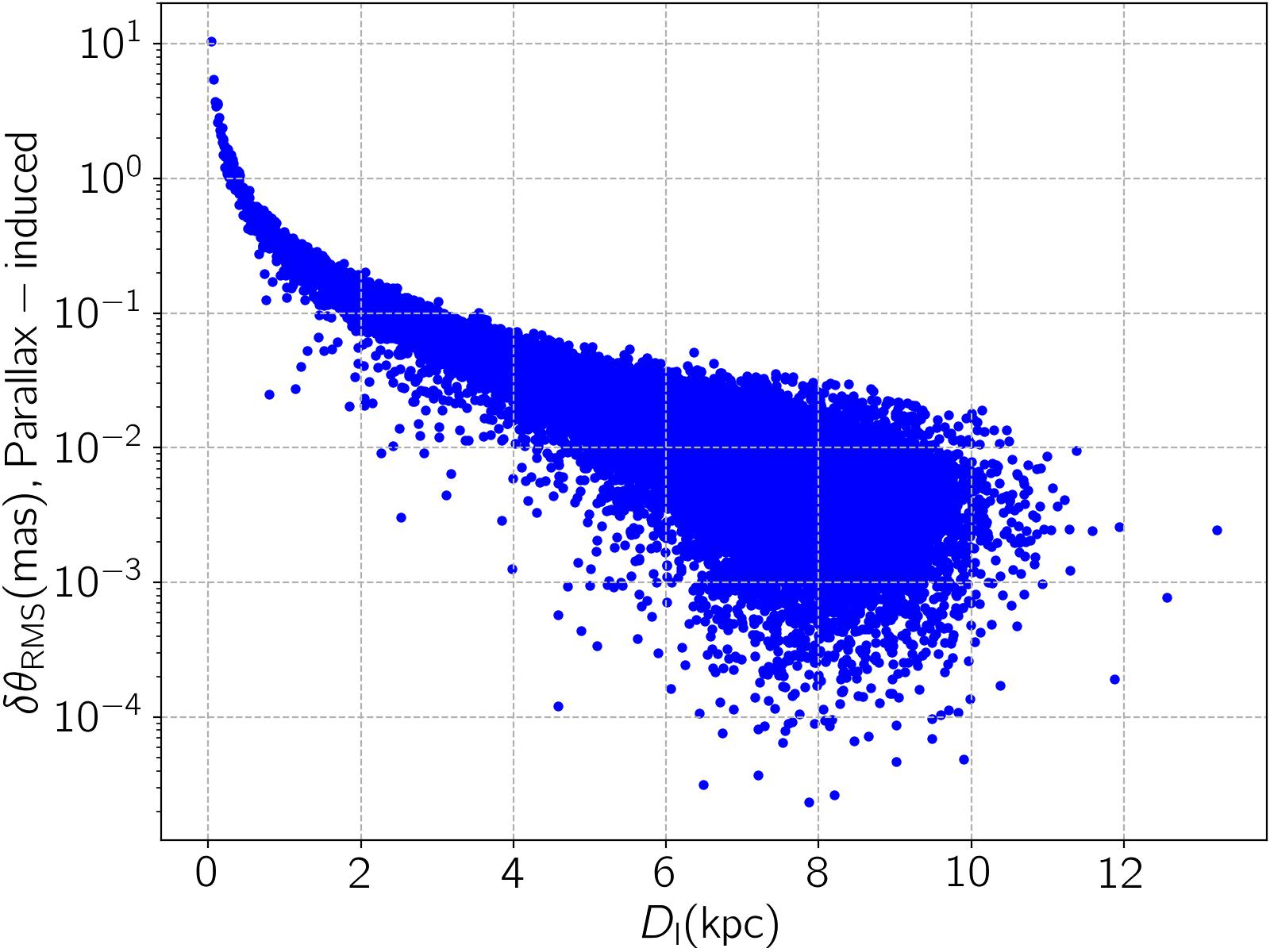}
\includegraphics[width=0.32\textwidth]{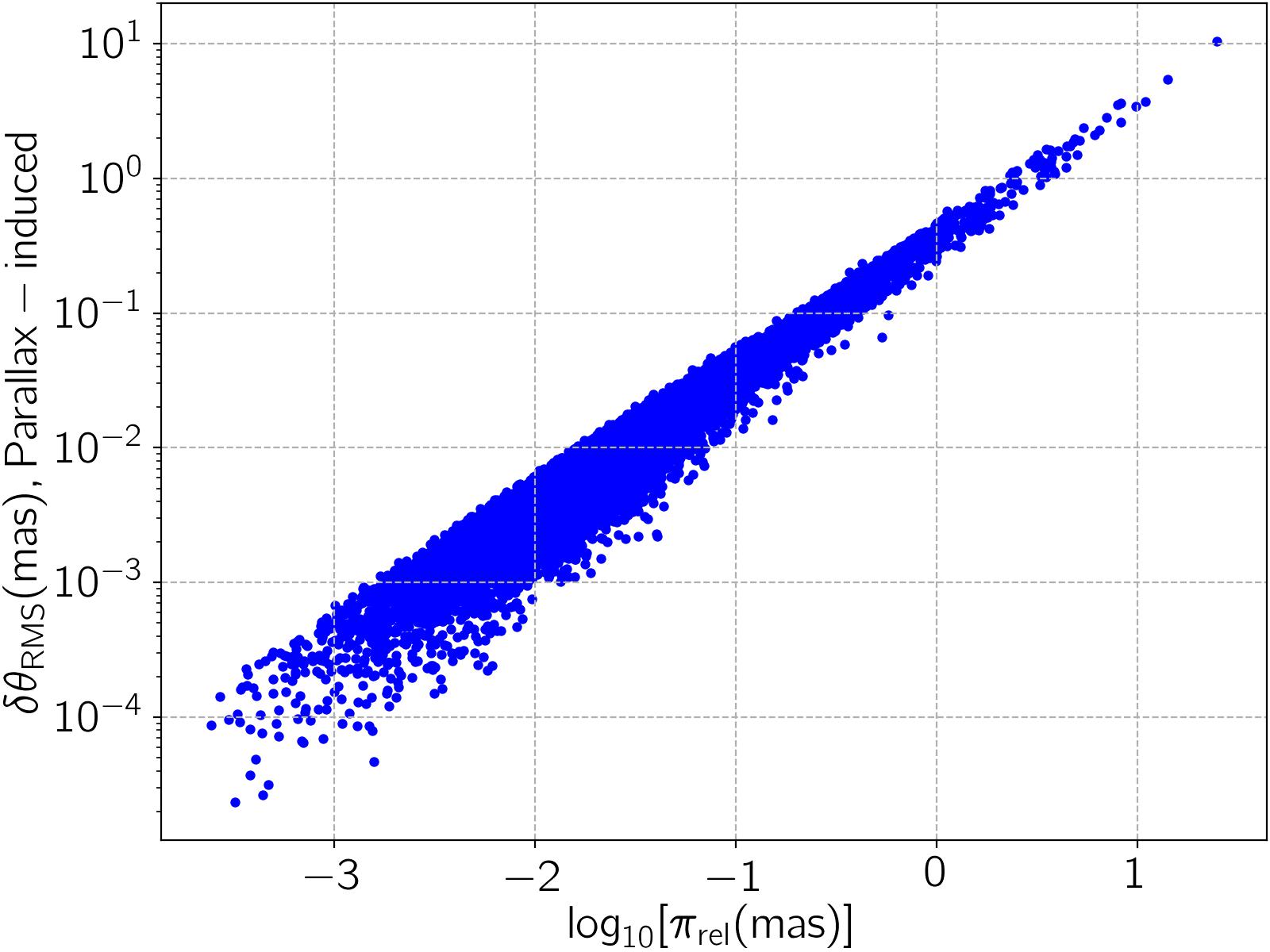}
\includegraphics[width=0.32\textwidth]{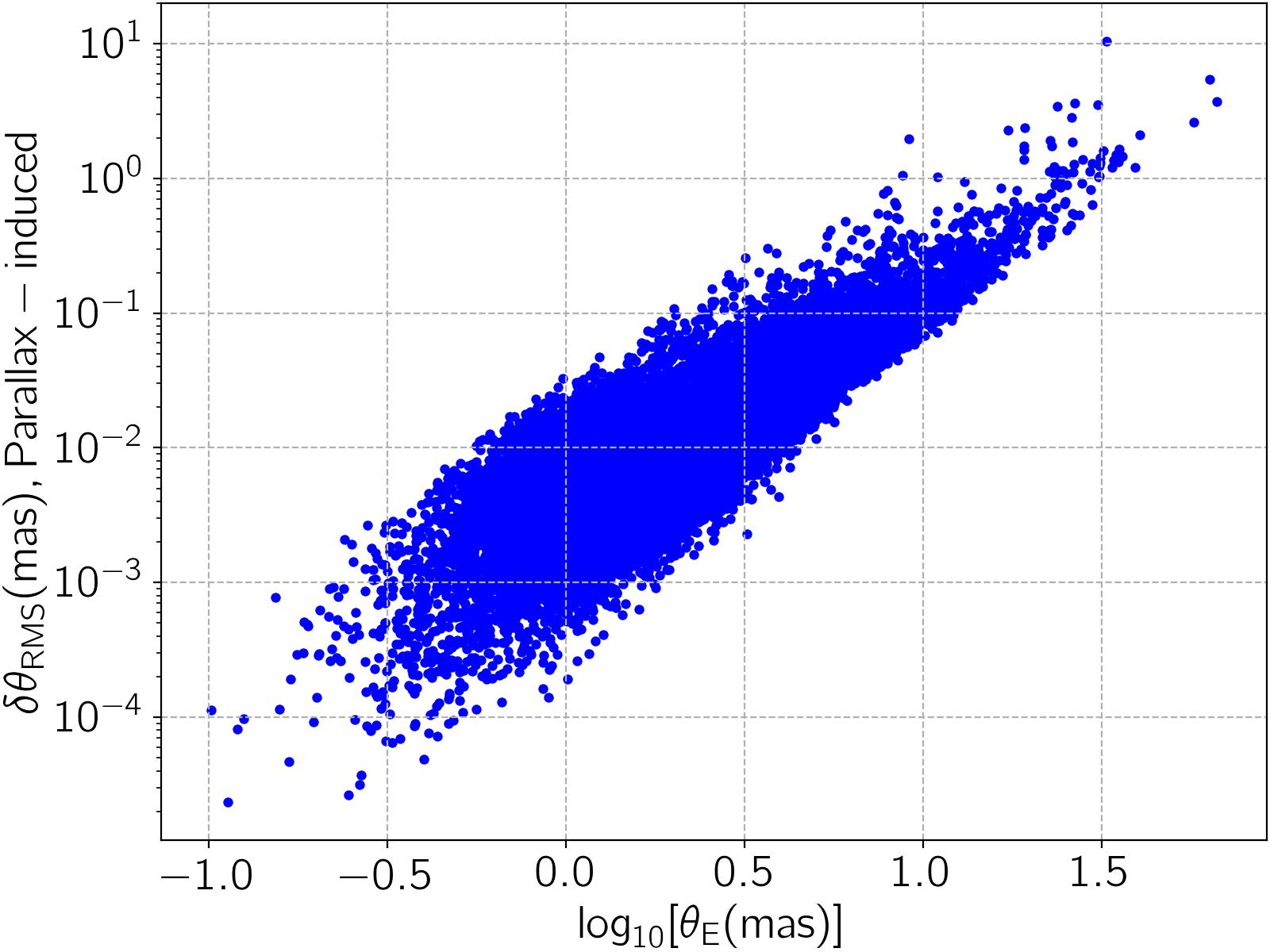}
\includegraphics[width=0.32\textwidth]{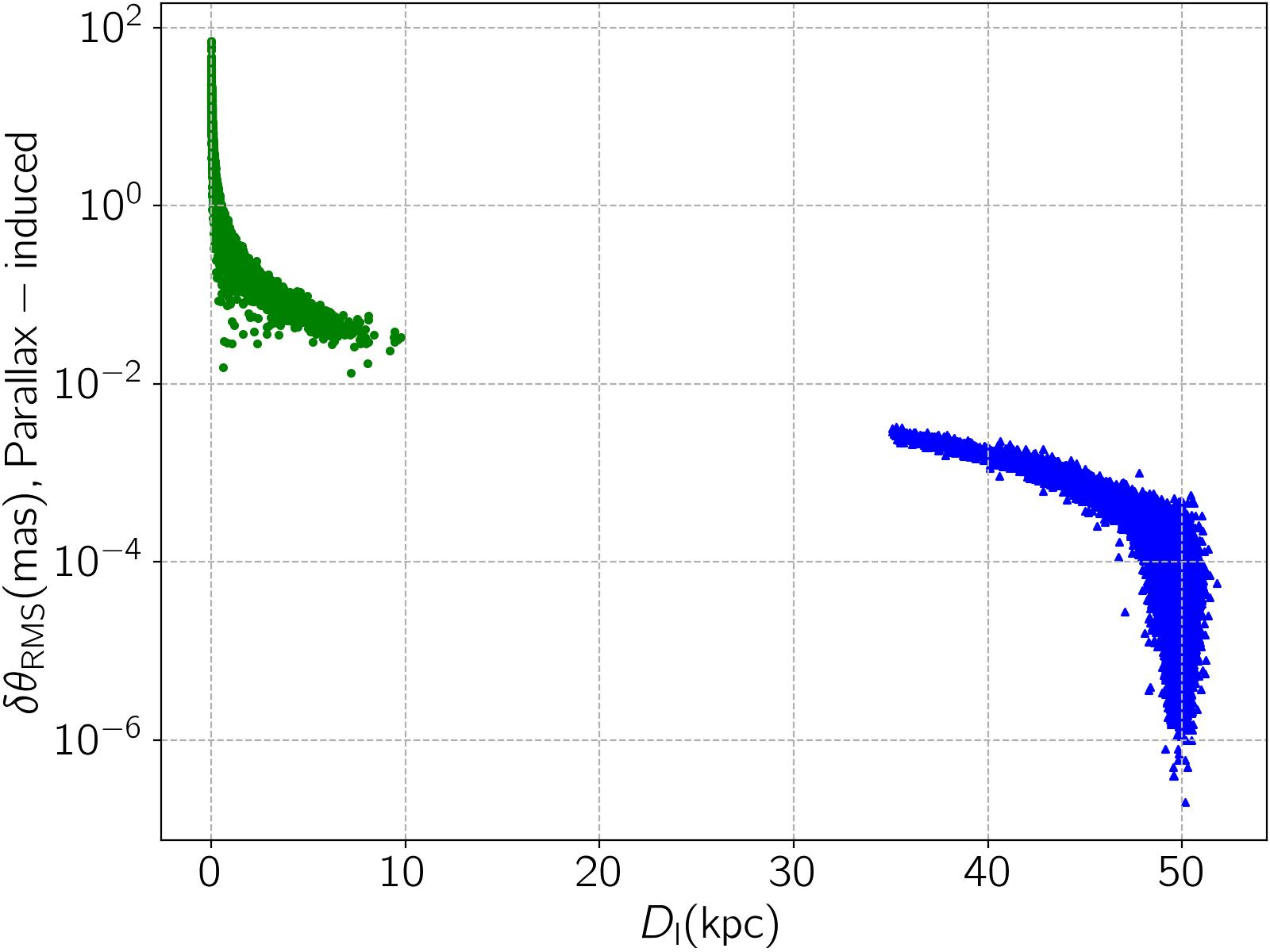}
\includegraphics[width=0.32\textwidth]{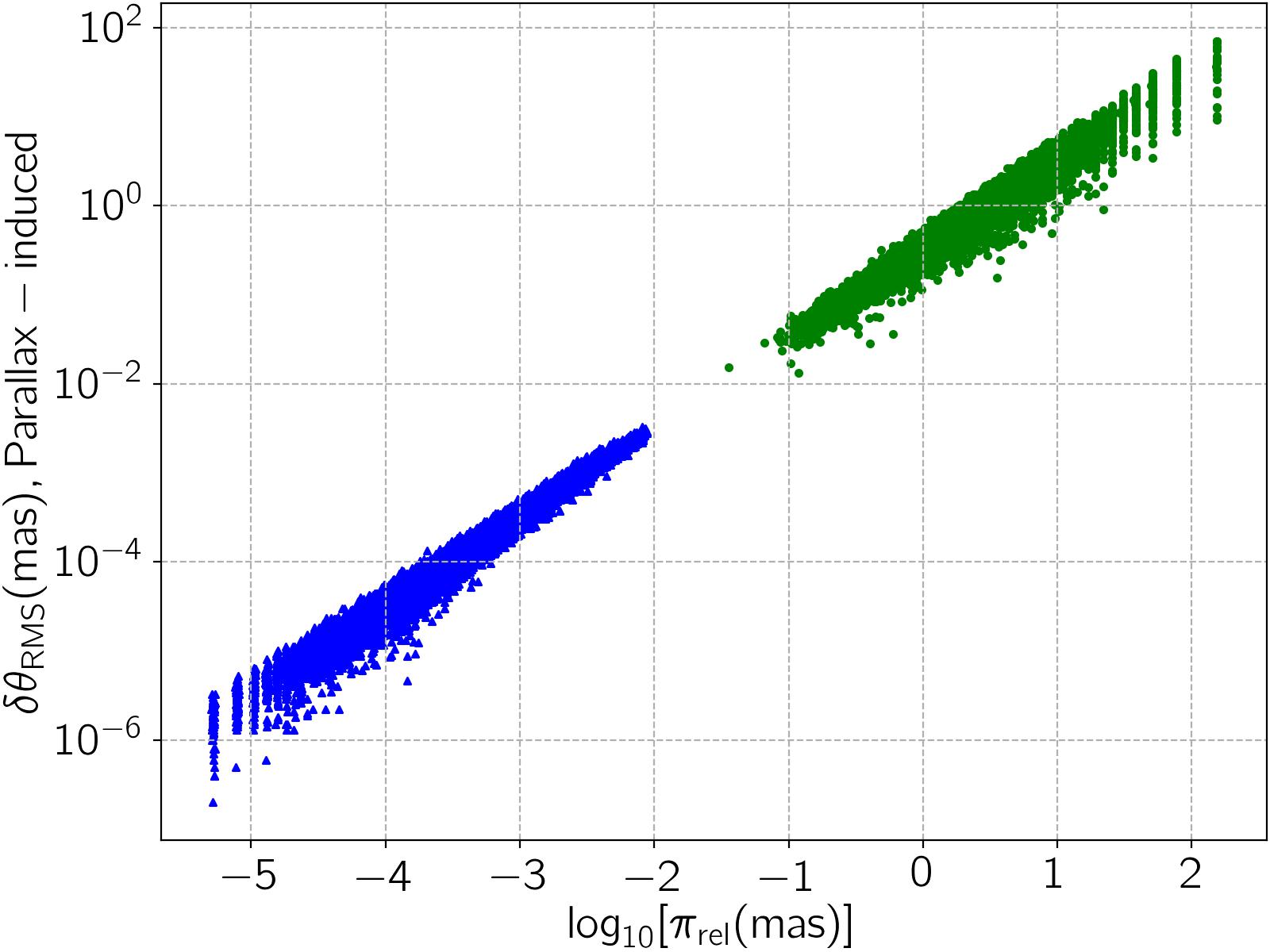}
\includegraphics[width=0.32\textwidth]{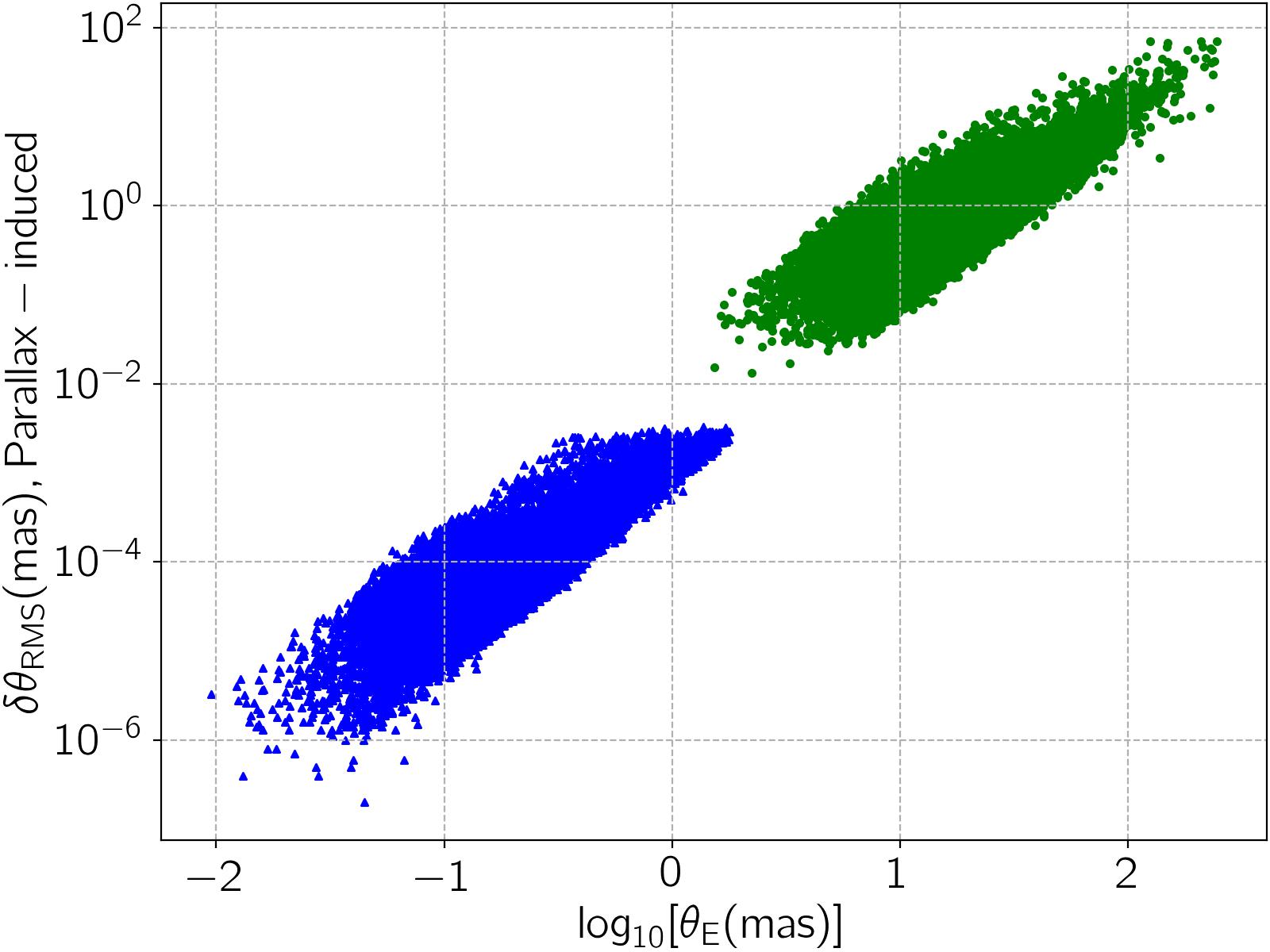}
\caption{Scatter plots of $\delta\theta_{\rm{RMS}}$ (given by Eq. \ref{deltat}) for a sample of astrometric microlensing events due to ISMBHs toward the GB (top panels), and LMC (bottom panels) versus $D_{\rm{l}}(\rm{kpc})$, $\log_{10}[\pi_{\rm{rel}} (\rm{mas})]$, and $\log_{10}[\theta_{\rm E}(\rm{mas})]$ (from left to right, respectively). In the bottom panels, self-lensing and halo-lensing events are specified with blue triangles and green circles, respectively.}\label{Fig_scatter}
\end{figure*}	 
\subsection{Astrometric microlensing toward the GB}\label{simul1}
Using the introduced formalism, we simulate possible astrometric microlensing events toward the GB, by considering the annual parallax effect (based on their distribution functions). 
We justify our simulation according to the \wfirst\ observing strategy. We assume the \wfirst\ orbit is a circle of radius $1.01$ au centered on the Sun (at the second Sun-Earth Lagrange point L2). However, in the simulation we increase the observing time to $19$ years, because the astrometric deflection tends to zero very slowly.

To simulate a microlensing event, we first choose the source distance from the observer, $D_{\rm s}$, using the projected mass density in each line of sight $dM(D_{\rm s},~l,~b)/dD_{\rm s}= d \Omega~D_{\rm s}^{2}~\big(\rho_{\rm b} +\rho_{\rm d} + \rho_{\rm h}\big)$. Here,  $(l,~b)$ are the Galactic longitude and latitude, respectively. $\rho_{\rm b}$, $\rho_{\rm d}$, and $\rho_{\rm h}$ are stellar densities of the Galactic bulge, disks, and halo, respectively. The physical properties of source stars (absolute magnitudes, mass, radius, type, ...) are determined according to the Galactic Besan\c{c}on model \footnote{\url{https://model.obs-besancon.fr/}}\citep{Robin2003, Robin2012}.  

The distance of the lens object from the observer depends on $D_{\rm s}$, and the given line of sight $(l,~b)$. We determine the lens distance from the observer using the microlensing event rate, i.e., $\Gamma \propto R_{\rm E}~\big(\rho_{\rm b}+\rho_{\rm d}+\rho_{\rm h}\big)~v_{\rm{rel}}$, where $R_{\rm E}=D_{\rm l}~\theta_{\rm E}$, $v_{\rm{rel}}=D_{\rm l}~\mu_{\rm{rel}}$, and $\mu_{\rm{rel}}$ is the size of the angular relative lens-source velocity (given by Eq. \ref{mul}). In order to simulate long-duration microlensing events due to ISMBHs, we select the lens mass uniformly from the range $M_{\rm l} \in [2,~50] M_{\odot}$ \citep{2022ApJSicilia}.

Three examples of astrometric deflections in the source trajectories are depicted in three top panels of Figure \ref{Fig_shift}. In these plots, solid black and dashed blue curves are astrometric deflections without and with considering the parallax effect. Dotted magenta curves represent the first term in Equation \ref{twoterm}. These magenta curves show that parallax-induced perturbations are mostly generated by the second term of Equation \ref{twoterm}. At the top of plots the relevant parameters are reported. These plots confirm the results in Subsection \ref{forma}.

Concerning the detectability of parallax-induced perturbations in astrometric deflections, we evaluate an statistical parameter. For each microlensing event, in the time interval $[-1.5~t_{\rm E},~1.5~t_{\rm E}]$, we numerically calculate the root mean square (RMS) of deviations in astrometric deflections, i.e., 
\begin{eqnarray}\label{deltat}
\delta\theta_{\rm{RMS}}= \sqrt{\left<\big(\delta\theta(t)-\delta\theta_{\odot}(t)\big)^{2}\right>_{t}},
\end{eqnarray}
where $\delta\theta_{\odot}(t)$ is the astrometric deflection without parallax effect. This parameter shows the scale of parallax-induced deviations in astrometric deflections.  
We simulate a large number of astrometric microlensing events due to massive lens objects toward the GB and extract the detectable events in the \wfirst\ observations. Then, we determine $\delta\theta_{\rm{RMS}}$ for each detectable event.

In three top panels of  Figure \ref{Fig_scatter}, we represent the scatter plots of $\delta\theta_{\rm{RMS}}$ versus $D_{\rm l}$, $\log_{10}\big[\pi_{\rm{rel}}(\rm{mas})\big]$, and $\log_{10}\big[\theta_{\rm E}(\rm{mas})\big]$, from left to right respectively. Accordingly, the parallax-induced perturbations in astrometric deflections are considerable when the lens is very close to the observer. Although the size of these perturbations does not depend on the lens mass, they are detected in the events due to more massive lenses with a higher probability. Because astrometric deflections are scaled with $\theta_{\rm E}\propto \sqrt{M_{\rm l}}$. Two last panels of Figure \ref{Fig_scatter} manifest that $\delta\theta_{\rm{RMS}} \propto \pi_{\rm{rel}} \propto \theta_{\rm E}^{2}$.

For the microlensing observations toward the Magellanic Clouds (MCs), the parallax perturbations in astrometric deflections should be even higher. Because the lens object can be in the Galactic halo ($D_{\rm l}\sim 2$ kpc) and the source stars are inside the Large Magellanic Cloud ($D_{\rm s}\sim 50$ kpc), which results a large relative parallax, i. e., $\pi_{\rm{rel}}\sim 0.48$ mas. We study this point in the next subsection.  

\subsection{Astrometric microlensing toward LMC}\label{simul2}

Microlensing observations toward LMC were first proposed by \citet{1986Paczynski} to determine the contribution of MACHOs in the Galactic halo. The upcoming LSST telescope will also monitor LMC with a $3$-day cadence in 6 filters, $ugrizy$, during its mission. This telescope will discern a considerable number of long-duration microlensing events, because of its long observing time (i.e., 10 years). Although, its long cadence is such that it is simply not going to be able to resolve the hours/day anomalies typical of planetary microlensing events. In this subsection, we simulate astrometric microlensing events toward LMC by considering the parallax effect. \\

The LMC celestial coordinate is $(\rm{RA},~\rm{DEC})= (80.9^{\circ}, ~-68.2)^{\circ}$, and its distance from the observer is $49.97$ kpc. For simulating potential microlensing events toward LMC, we include stellar spacial distributions due to the LMC's disk, bulge, and halo, as given by \citet{2000Gyuk}, and rewritten in Appendix (A) of \citet{2021sajadianFFP}. We assume the photometric properties of the LMC stars are the same as those in our galaxy. The extinction toward LMC is ignorable except its central part \citep[see, e.g., ][]{2008Dobashi}. For the central part of LMC, we consider the $V$-band extinction in the range of $A_{V} \in [1.6,~2.1]$ mag uniformly. We simulate microlensing events toward a square of angular side $4$ degree centered on the LMC's center.   

Three examples of simulated astrometric deflections in microlensing events toward LMC are represented in bottom panels of Figure \ref{Fig_shift}. The last one is a self-lensing event (its lens object is inside LMC), and the two others are halo-lensing ones (their lens objects are in the Galactic halo). Accordingly, the parallax effect is realizable in astrometric deflections of halo-lensing events made by massive lens objects.   

In bottom panels of Figure \ref{Fig_scatter}, we show the scatter plots of $\delta\theta_{\rm{RMS}}$ versus the lens distance, $\log_{10}\left[\pi_{\rm{rel}}(\rm{mas})\right]$, and $\log_{10}\left[\theta_{\rm E}(\rm{mas})\right]$. In these plots, halo-lensing events are specified with green circles and self-lensing events are denoted with blue triangles. On average, relative parallax amplitude ($\pi_{\rm{rel}}$), angular Einstein radius, and as a result $\delta\theta_{\rm{RMS}}$ in halo-lensing events are larger than those due to self-lensing events by more than two orders of magnitude. 

Also, by changing the observational direction from the GB to LMC, $\delta\theta_{\rm{RMS}}$ enhances more than one order of magnitude. The astrometric accuracy of ELT\footnote{\url{https://elt.eso.org/}}\citep{ 2011PASPELT,ELTpaper}, which is under construction, for a bright star with a $K$-band apparent magnitude $18-19$ mag reaches $50~\mu$as. Hence, this telescope by following up long-duration events toward LMC can detect not only astrometric deflections in source trajectories, but also their parallax-induced deviations. Measuring $\theta_{\rm E}$, and $\pi_{\rm{rel}}$ will specify the lens mass, and its distance uniquely. We study this point in the next section quantitatively.

\section{Simulations of astrometric microlensing by \wfirst\ and LSST}\label{mcmc}
In previous section, we found that the parallax effect could make considerable deviations in lensing-induced astrometric deflections due to massive and close lens objects. Here, we aim to evaluate detectability of these parallax-induced deviations and their characterizations.

Extracting parallax amplitudes depend on the observing photometric and astrometric accuracies, observing cadence, and the observing time interval. Therefore, evaluating efficiency for discerning and characterizing parallax amplitudes in astrometric deflections or microlensing light curves needs comprehensive simulations of these events by generating synthetic data points based on real observations.  

For more realistic simulations, we apply observing strategies due to upcoming microlensing surveys. We consider three strategies which are summarized in Table \ref{tab1}, and explained in the following
\begin{itemize}[leftmargin=2.0mm]
\item {\bf A: }Survey observations with the \wfirst\ telescope toward the GB during six 62-day seasons (its total observing time is $5$ years) with a $15$-min cadence. We also consider some extra observations (one hour observation every ten days) during the \wfirst\ large seasonal gap and when the bulge is visible for this telescope. The same observing strategy was introduced to study detecting ISMBHs by the \wfirst\ telescope in \citet{2023sajadiansahu}.   

\item {\bf B: }Survey observations with the \wfirst\ telescope and follow-up observations with the ELT telescope in $K$-band toward the Galactic bulge. The ELT telescope will start observations for each event when $A>1.34$. We assume this telescope will take one data point every $10$ days during its observing seasons.    

\item {\bf C: }Survey observations with the Vera C. Rubin Observatory's Legacy Survey of Space and Time (LSST) telescope toward the LMC, and follow-up observations with the ELT telescope. The LSST cadence is planned to be $3$ days, and its mission will take $10$ years.  
\end{itemize}
We tune our simulations based on these observing strategies. In the following, we explain the details of these Monte Carlo simulations, and the results.

\begin{figure*}
\centering
\includegraphics[width=0.49\textwidth]{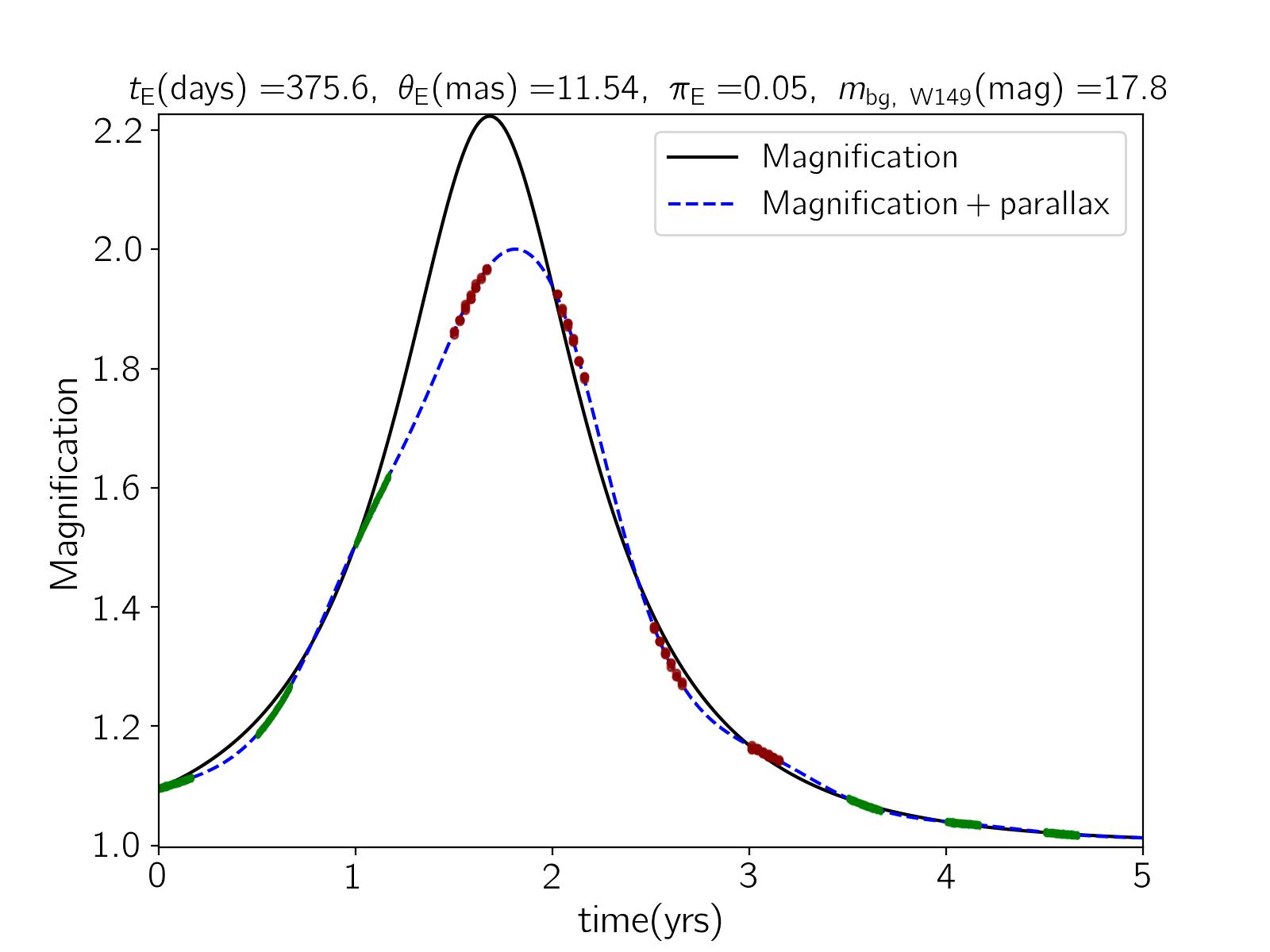}
\includegraphics[width=0.49\textwidth]{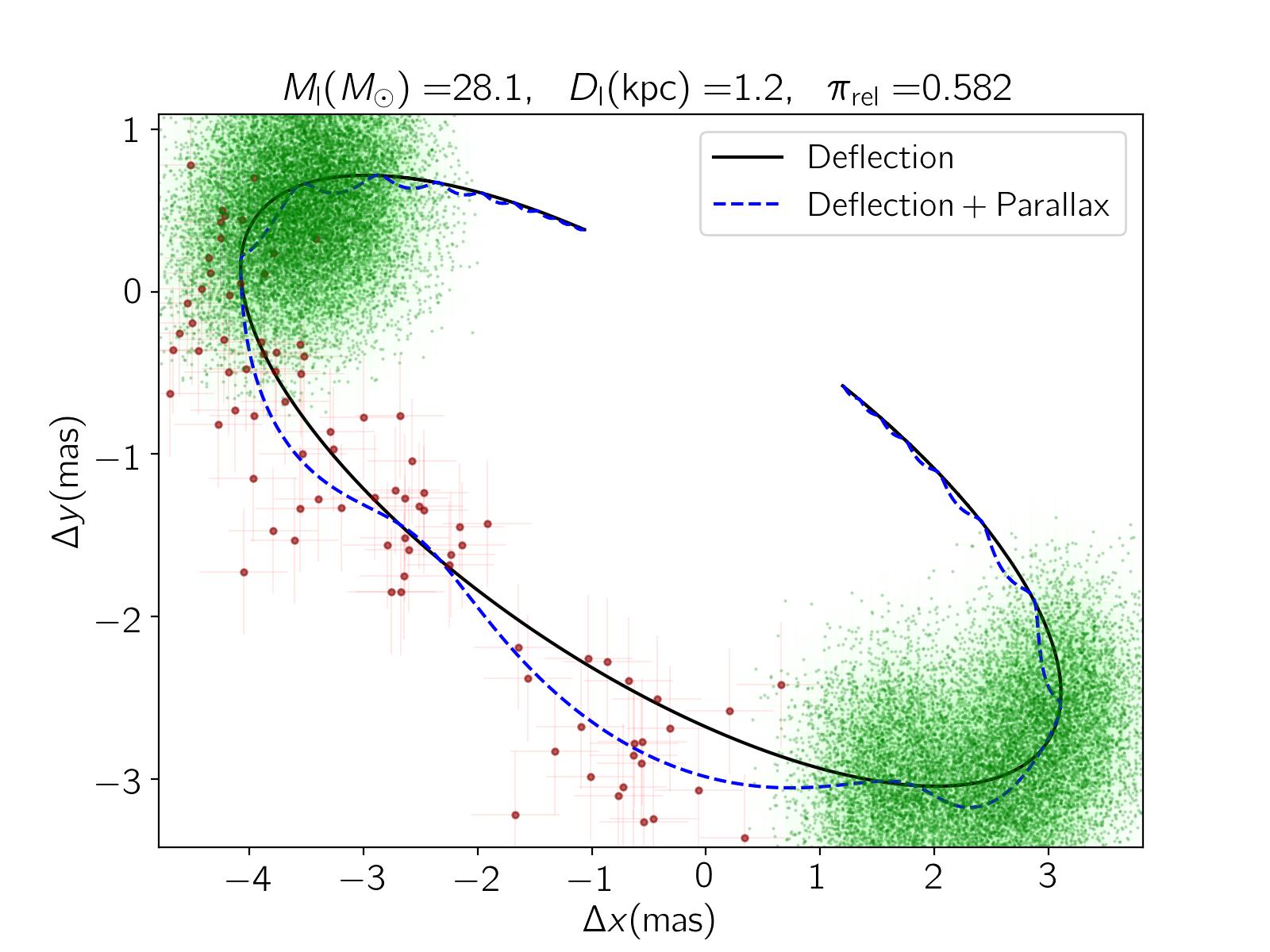}
\includegraphics[width=0.49\textwidth]{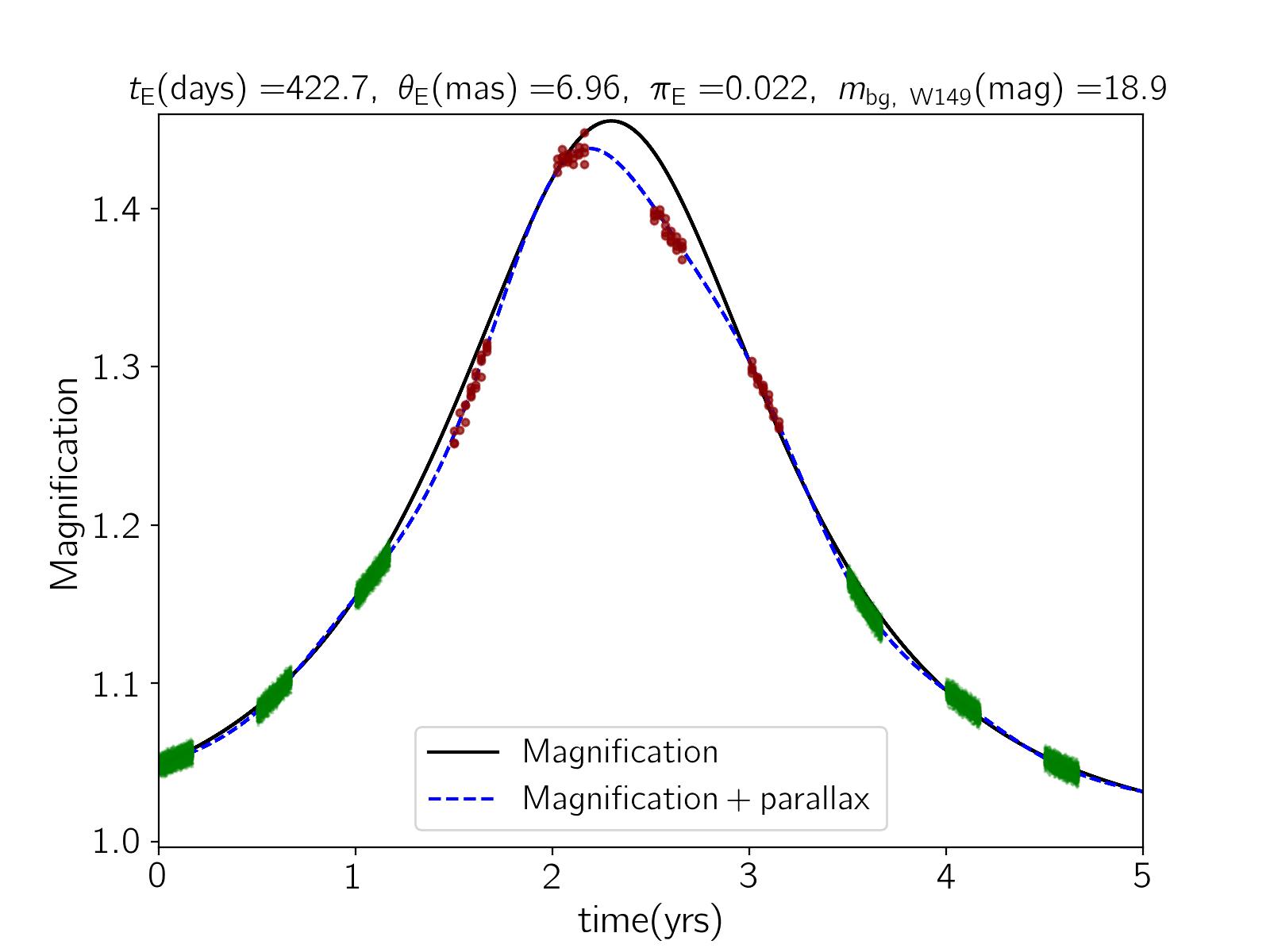}
\includegraphics[width=0.49\textwidth]{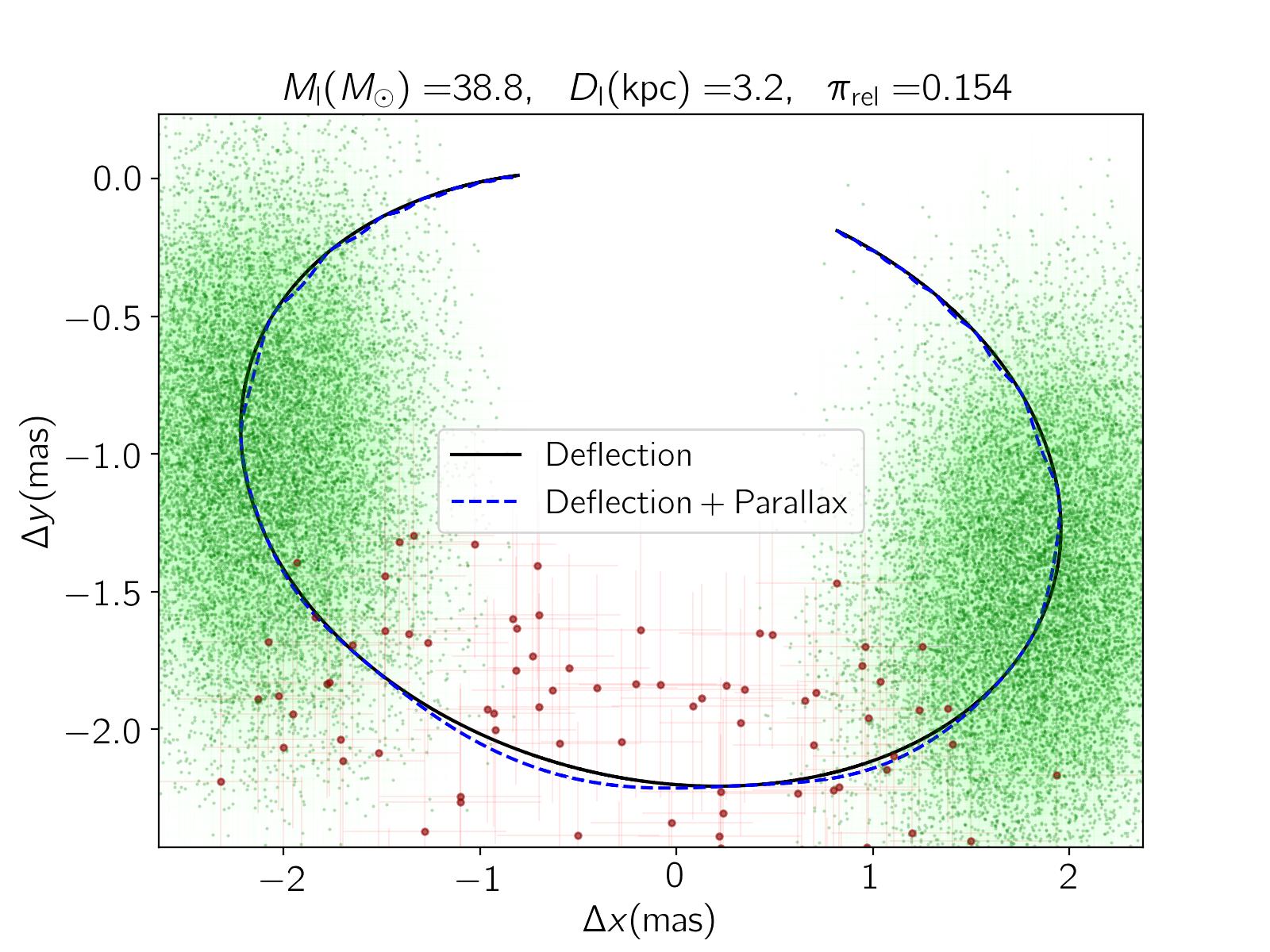}
\caption{Two examples of simulated microlensing events which are detectable with the \wfirst\ telescope. For each event its light curve and astrometric deflection are represented alongside (left and right panels). The synthetic green and dark red data points are taken by the \wfirst\ telescope during its observing seasons, and its large seasonal gap, respectively. Their parameters can be found at the top of plots.}\label{Roman1}
\end{figure*}

\begin{deluxetable}{cccccc}
\tablecolumns{6}
\centering
\tablewidth{0.48\textwidth}\tabletypesize\footnotesize
\tablecaption{The descriptions of Monte Carlo simulations (A), (B), and (C) which are explained in Section \ref{mcmc}. \label{tab1}}
\tablehead{\colhead{Simulation}&\colhead{Observations}&\colhead{Telescope}& \colhead{$T_{\rm{obs}}(\rm{yrs})$}&\colhead{Cadence}& \colhead{Filter}}
\startdata	
(A) & Survey &   \wfirst\  & 5 &  $15$min  & W149 \\
	   & Follow-up &  \wfirst\ & 5 & $10$ days & W149\\
\tableline
(B) & Survey &   \wfirst\  & 5 &  $15$min  & W149 \\
 	   & Follow-up &   ELT & 5 & $10$ days & $K$\\
\tableline
(C) & Survey &   LSST  & 10 &  $3$ days  & $ugrizy$ \\
   	  & Follow-up &  ELT & 10 & $10$ days &  $K$\\
\enddata
\end{deluxetable}

\begin{deluxetable*}{c c c c c c c c c}
\tablecolumns{9}
\centering
\tablewidth{0.99\textwidth}\tabletypesize\footnotesize
\tablecaption{The results of Monte Carlo simulations (A), (B), and (C) which are explained in Tab \ref{tab1}.\label{tab2}}
\tablehead{\colhead{$\rm{Simulation}$}&\colhead{$\epsilon_{\theta_{\rm E}}$}&\colhead{$\epsilon_{\pi_{\rm E}}$}&\colhead{$\epsilon_{\pi_{\rm E},~\rm{Phot}}$} &\colhead{$\epsilon_{\pi_{\rm E},~\rm{Ast}}$} & \colhead{$\epsilon_{M_{\rm l}}$} & \colhead{$\epsilon_{D_{\rm l}}$}&\colhead{$\epsilon_{t_{\rm E}}$}&  $\epsilon$\\
&$[\%]$&$[\%]$&$[\%]$&$[\%]$ & $[\%]$ & $[\%]$& $[\%]$ & $[\%]$}
\startdata  
\multicolumn{9}{c}{Detection Threshold = $1\%$}\\
(A)&$70.65$ & $7.80$ & $7.72$ & $0.36$ & $5.48$ & $21.76$ & $29.73$ & $5.71$ \\
(B)& $76.71$ & $11.35$ & $11.30$ & $0.78$ & $9.18$ & $29.87$ & $49.78$ & $9.75$\\
(C)&$62.14$ & $25.92$ & $8.38$ & $25.08$ & $8.03$ & $18.69$ & $20.88$ & $10.20$ \\
\tableline
\multicolumn{9}{c}{Detection Threshold = $4\%$}\\
(A)& $97.47$ & $21.94$ & $21.68$ & $2.16$ & $21.05$ & $46.67$ & $58.40$ & $20.94$\\
(B) &$98.25$ & $29.18$ & $29.09$ & $3.76$ & $28.41$ & $59.57$ & $85.78$ & $28.83$ \\
(C) & $77.12$ & $42.00$ & $22.98$ & $41.09$ & $22.80$ & $47.67$ & $54.64$ & $27.44$\\
\tableline
\multicolumn{9}{c}{Detection Threshold = $7\%$}\\
(A)&$99.36$ & $29.81$ & $29.44$ & $4.05$ & $29.19$ & $57.57$ & $69.29$ & $28.93$\\
(B)&$99.61$ & $38.61$ & $38.46$ & $6.55$ & $38.18$ & $71.18$ & $93.34$ & $38.46$ \\
(C)& $82.78$ & $47.27$ & $29.93$ & $46.10$ & $29.78$ & $59.76$ & $67.97$ & $35.04$\\
\enddata
\tablecomments{$\epsilon_{a}$ refers to the probability of measuring the parameter $a$ so that its relative error be less than the given threshold. The indices 'Phot' and 'Ast' in the probability of measuring parallax amplitudes  refer to infer parallax from photometric and astrometric data with a given detection threshold, respectively.}
\end{deluxetable*}

{\bf A: Survey Microlensing with \textit{Roman}:~}We first tune our simulations for potential survey observations toward the Galactic bulge with the \wfirst\ telescope. We have done similar simulations in \citet{2023sajadiansahu}. In that paper, we showed that a small number of additional observations, i.e., one hour of observations every 10 days, when the Galactic bulge is observable during the large seasonal gap will improve the \wfirst\ efficiency for detecting and characterizing ISMBHs. In the simulation (A), we assume these extra and sparse data points are taken with \wfirst.

In the simulation, we take the time of the closest approach $t_{0}$ uniformly in the range $[0,~5]$ years. The lens impact parameter is also chosen smoothly from the range $[0,~1]$. The finite source effect for microlensing events due to massive lens objects is ignorable because the normalized source radius projected on the lens plane is $\rho_{\star} \propto 1/\sqrt{M_{\rm l}}$, and very small. We consider the parallax effect while calculating light curves, and astrometric deflection, to compare the errors while extracting the parallax amplitude from each of them. Since the \wfirst\ telescope will orbit the Sun from the second Sun-Earth Lagrange point (L2), the radius of its orbit is $1.01$ au.

In Monte Carlo simulations from microlensing events detectable by \wfirst\ we remove events with time scale $t_{\rm E}>2000$ days. Because in most of these events magnification factors do not reach to the baseline during the \wfirst\ $5$-year mission.
   
For calculating the blending effect in all Monte Carlo simulations, we first calculate the average number of background stars inside a stellar PSF as $$\overline{N_{\star}}= \Omega_{\rm{PSF}}~f_{\rm b}~\int_{0}^{D_{\rm s}}\Big( \frac{\rho_{\rm b}(x) }{\overline{M_{\rm b}}} +  \frac{\rho_{\rm d}(x)}{\overline{M_{\rm d}}}+\frac{\rho_{\rm h}(x)}{\overline{M_{\rm h}}}\Big) x^{2}dx,$$
where, $f_{\rm b}=2/3$ is the binary fraction, $\overline{M_{\rm b}}$, $\overline{M_{\rm d}}$, and $\overline{M_{\rm h}}$ are the average masses of stars in the Galactic bulge, disk, and halo, respectively. $\Omega_{\rm{PSF}}=\pi\Big(\rm{FWHM}/2\Big)^{2}$ is the area of a typical stellar PSF in the \wfirst\ observations. We determine the number of stars using $N_{\star}=\overline{N_{\star}}+\sigma_{\rm N}$, where $\sigma_{\rm N}$ is chosen using a normal distribution $\mathcal{N}(0,~\sqrt{\overline{N_{\star}}})$. A common value for FWHM is $3\times$pixel size, i.e., $0.33$ arcsec for the \wfirst\ observation.  \\ 

The \wfirst\ photometry accuracy $\sigma_{\rm m}$ is a function of stellar apparent magnitude in the W149 filter and was given in Fig. (4) of \citet{2019Penny}. Stellar absolute magnitudes in W149 are estimated by $M_{\rm{W149}}\simeq (M_{H}+M_{J}+M_{K})/3$, where $M_{H}$, $M_{J}$, and $M_{K}$ represent the stellar absolute magnitudes in the standard filters $H$, $J$, and $K$, respectively \citep[see, e.g., ][]{2021sajadianHZP}. We note that the Galactic Becan\c{c}on model gives the stellar absolute magnitudes in the standard filters $BVRIK$. We specify the absolute magnitudes in $H$-, and $J$-bands using Dartmouth Isochrones\footnote{http://stellar.dartmouth.edu/}\citep{Dartmouth2008,DartMouth2011}. 

\noindent We use the 3D extinction map offered by \citet{Marshal2006}, and use the following transformation relations to specify extinctions in other bands: $A_{V}=8.47 A_{K}=8.21A_{K_{\rm s}}=1.67A_{I}=5.43A_{H}=4.44 A_{\rm{W149}}$ \citep{Cardelli1989}. We determine the astrometric accuracy of \wfirst\ according to stellar apparent magnitude, and using \textit{Jitter} simulations done by S. C. Novati. We assume the probability of regular observations with \wfirst\ is $90\%$ during its observing seasons.  

In Figure \ref{Roman1}, two simulated astrometric microlensing events detectable by \wfirst\ are represented. The sparse data points during its large seasonal gap are shown with dark red colour, and the \wfirst\ data during its observing seasons are depicted with green colour. In these figures the sparse data points rather help to discern astrometric deflections, and as a result the Einstein angular radius. The parallax amplitude can be extracted from the \wfirst\ photometric data.

\begin{figure}
\centering
\includegraphics[width=0.49\textwidth]{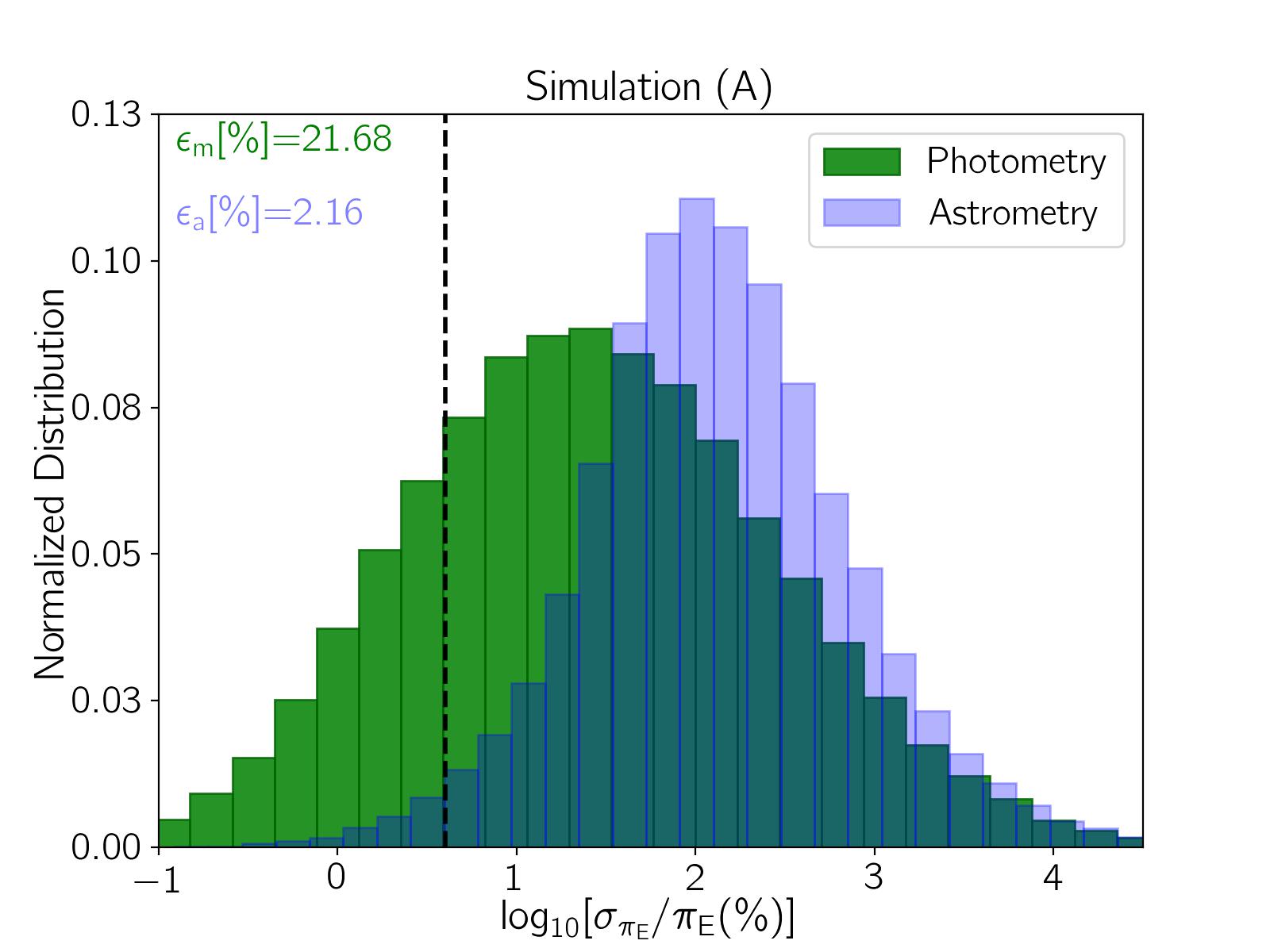}
\includegraphics[width=0.49\textwidth]{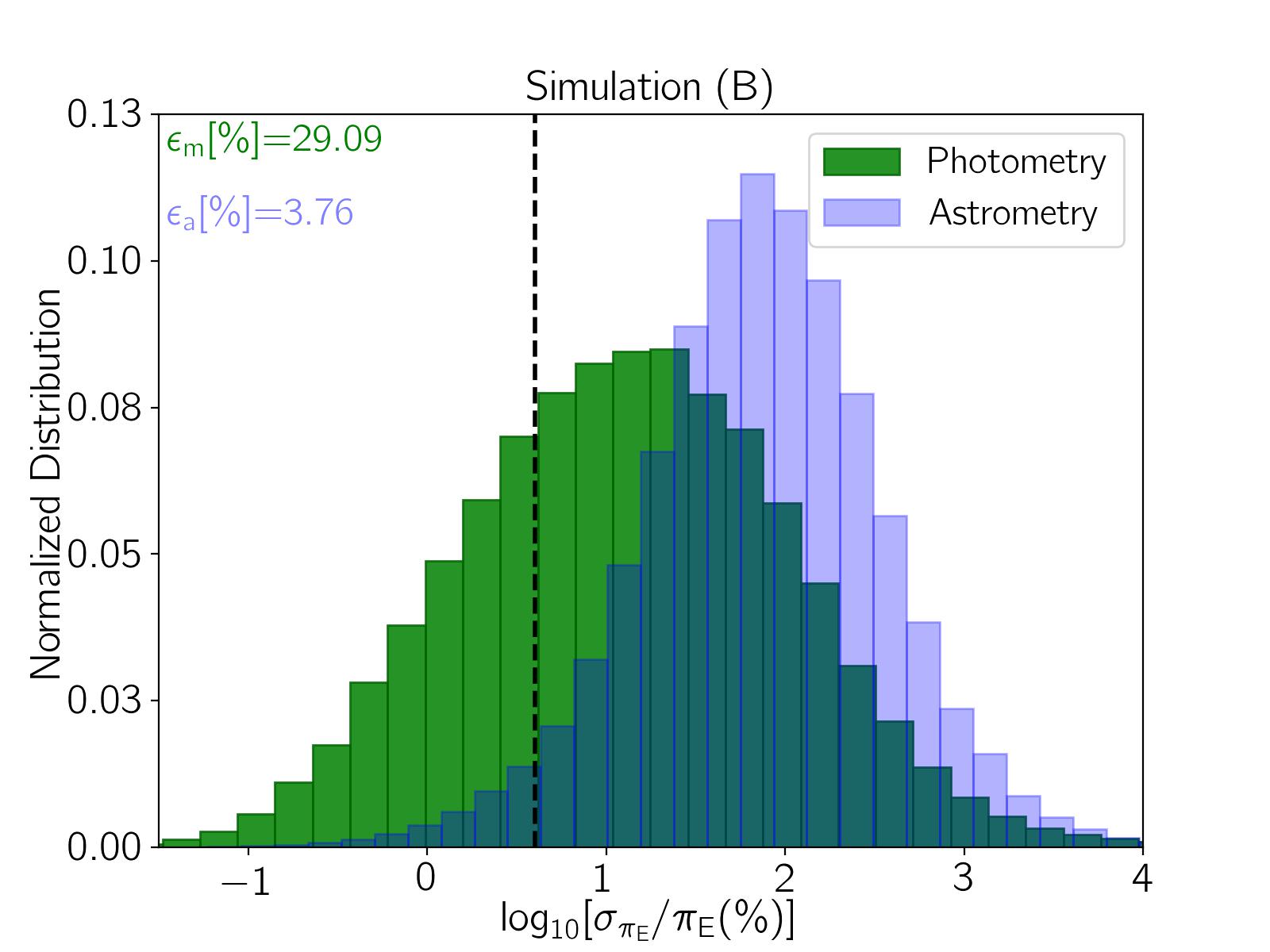}
\includegraphics[width=0.49\textwidth]{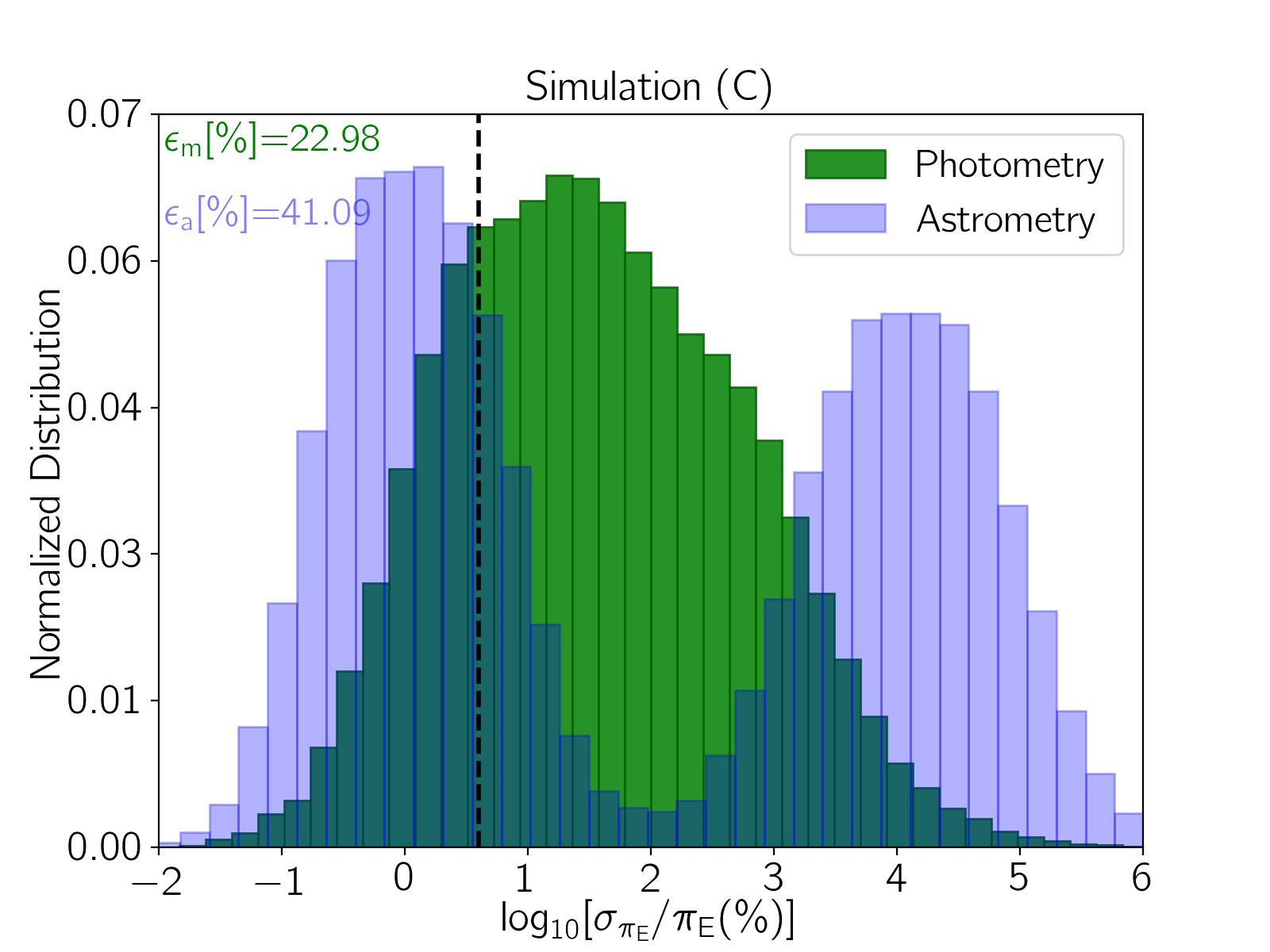}
\caption{The normalized distributions of relative errors in the parallax amplitudes, as measured through photometric (green) and astrometric observations (purple ones) based on simulations (A), (B), and (C), from top to bottom, respectively. The dashed black lines determine the threshold amount $\sigma_{\pi_{\rm E}}/\pi_{\rm E}=4\%$. The fractions of events with the relative parallax errors less than $4\%$ are mentioned inside the panels.}\label{sigmapi}
\end{figure}

After making a big ensemble of these events, we extract the detectable events in the \wfirst\ observations. We have two criteria for detectability as follows. (i) $\Delta
 \chi^{2}=\left| \chi^{2}_{\rm{real}}-\chi^{2}_{\rm{base}}\right|>800$, where $\chi^{2}_{\rm{real}}$, and $\chi^{2}_{\rm{base}}$ are $\chi^{2}$s from fitting the real model and the baseline, respectively. (ii) Three data point should be above the baseline by at least $4\sigma$, where $\sigma$ is the photometric accuracy. 
 
\noindent We numerically calculate the photometric and astrometric Fisher matrices separately, i.e., $\mathcal{A}$, and $\mathcal{B}$ for each detectable event. In this regard, observable parameters that affect on magnification curves and astrometric deflections are $t_{0},~u_{0},~t_{\rm E},~\xi,~f_{\rm b},~m_{\rm{base}},~\pi_{\rm E}$, and $\theta_{\rm E},~\pi_{\rm E}, ~\boldsymbol{\mu}_{\rm s}$, respectively. We embed $\pi_{\rm E}$ into both lists to compare the photometric and astrometric errors while extracting the parallax amplitude.

\noindent In fact, real astronomical data points determine the projected source trajectory on the sky plane, which is 
\begin{eqnarray}
\boldsymbol{\theta}_{\rm s}(t)=  \boldsymbol{u_{0}}~\theta_{\rm E} + \boldsymbol{\mu}_{\rm s} (t-t_{0})-\pi_{\rm s}\boldsymbol{\Delta}_{\rm o,~n}(t) +\boldsymbol{\delta \theta}(t), 
\end{eqnarray} 
where, $\boldsymbol{\mu}_{\rm s}$  is the angular velocity of source star, $\pi_{\rm s}= 1.01~\rm{au}/D_{\rm s}$ is the so-called source parallax amplitude when the observer is \wfirst.\ The source parallax amplitude ($\pi_{\rm s}$) is very small and ignorable. Hence, the parallax-induced perturbations mostly alter the last term (i.e., the astrometric deflection).   
Throughout the paper and in Figures \ref{Fig_shift}, \ref{Roman1}, \ref{Roman2}, and \ref{LSST}, we only show the last term (the astrometric deflection) so that the parallax-induced deviations get highlighted. However, while calculating the astrometric Fisher matrix ($\mathcal{B}$) we numerically calculate derivatives of the source trajectory $\boldsymbol{\theta}_{\rm s}(t)$ with respect to its parameters.

\noindent  The results from this simulation are reported in Table \ref{tab2}. In this table, $\epsilon_{a}$ is the probability (or efficiency) of measuring the parameter $a$ with the relative error less than the given thresholds (i.e., $1\%$, $4\%$, and $7\%$). The indices  'Phot' and 'Ast' for $\epsilon_{\pi_{\rm E}}$  refer to measuring parallax amplitudes from photometric and astrometric data, respectively.  The last column is the probability of simultaneously measuring three parameters $\theta_{\rm E}$, $\pi_{\rm E}$, and $t_{\rm E}$ with the relative errors less than the given threshold. These parameters uniquely offer the lens mass, the lens distance, and the lens-source relative velocity. The relations to calculate errors can be found in \citet{2023sajadiansahu}, and we do not repeat them here.  

The normalized distributions of relative errors in the parallax amplitude, $\sigma_{\pi_{\rm E}}/\pi_{\rm E}$, extracted from Covariance matrices $\mathcal{A}^{-1}$ (based on photometric data), and $\mathcal{B}^{-1}$ (based on astrometric data) are depicted in the first panel of Figure \ref{sigmapi} with green and purple colours, respectively.

Accordingly, in long-duration microlensing events toward the Galactic bulge detectable by the \wfirst\ telescope, the parallax can be took out from light curves much more than from astrometric deflections ($10$ times). This point can be discovered in Figures \ref{Roman1}.

\begin{figure*}
	\centering
	\includegraphics[width=0.32\textwidth]{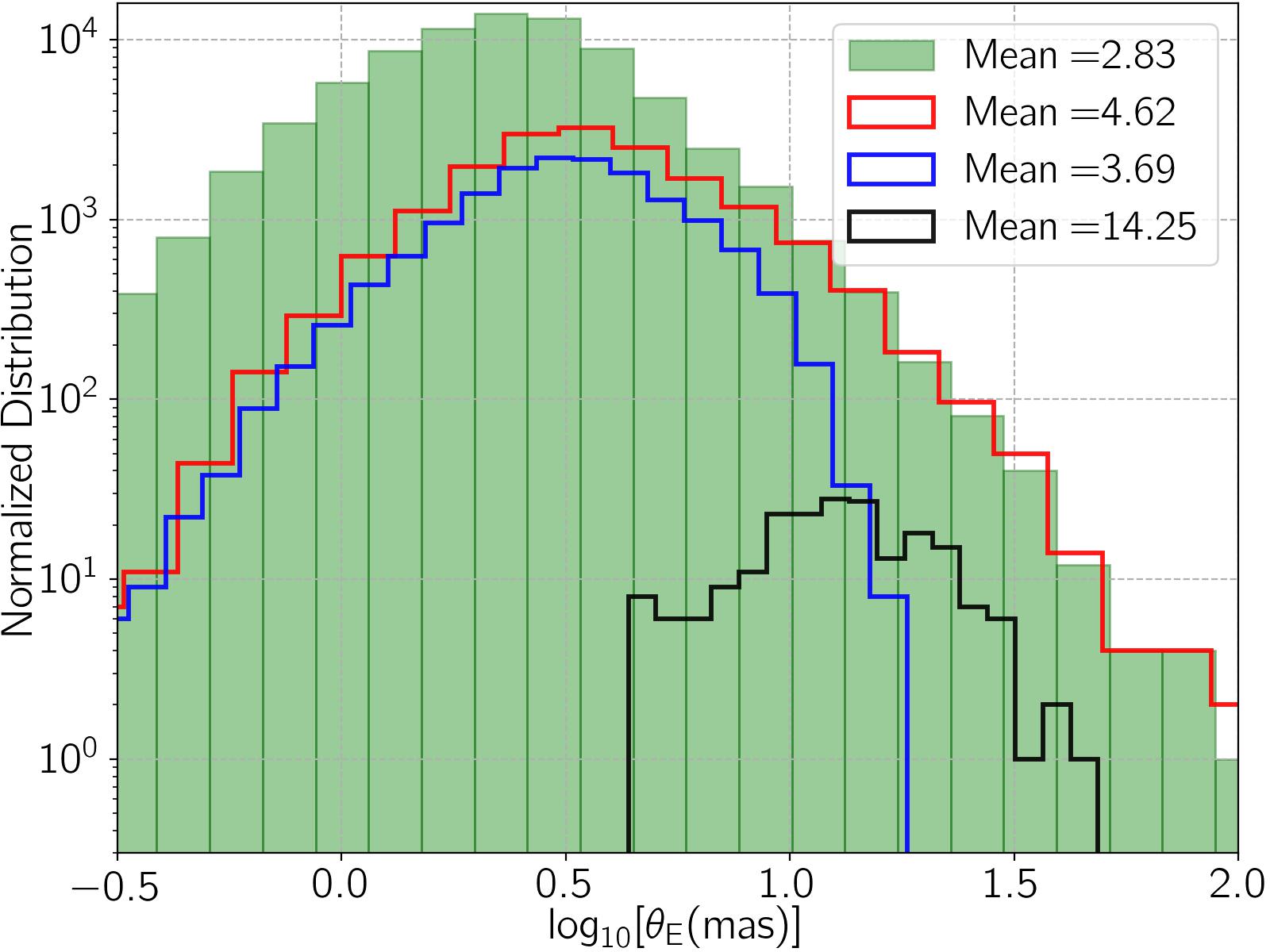}
	\includegraphics[width=0.32\textwidth]{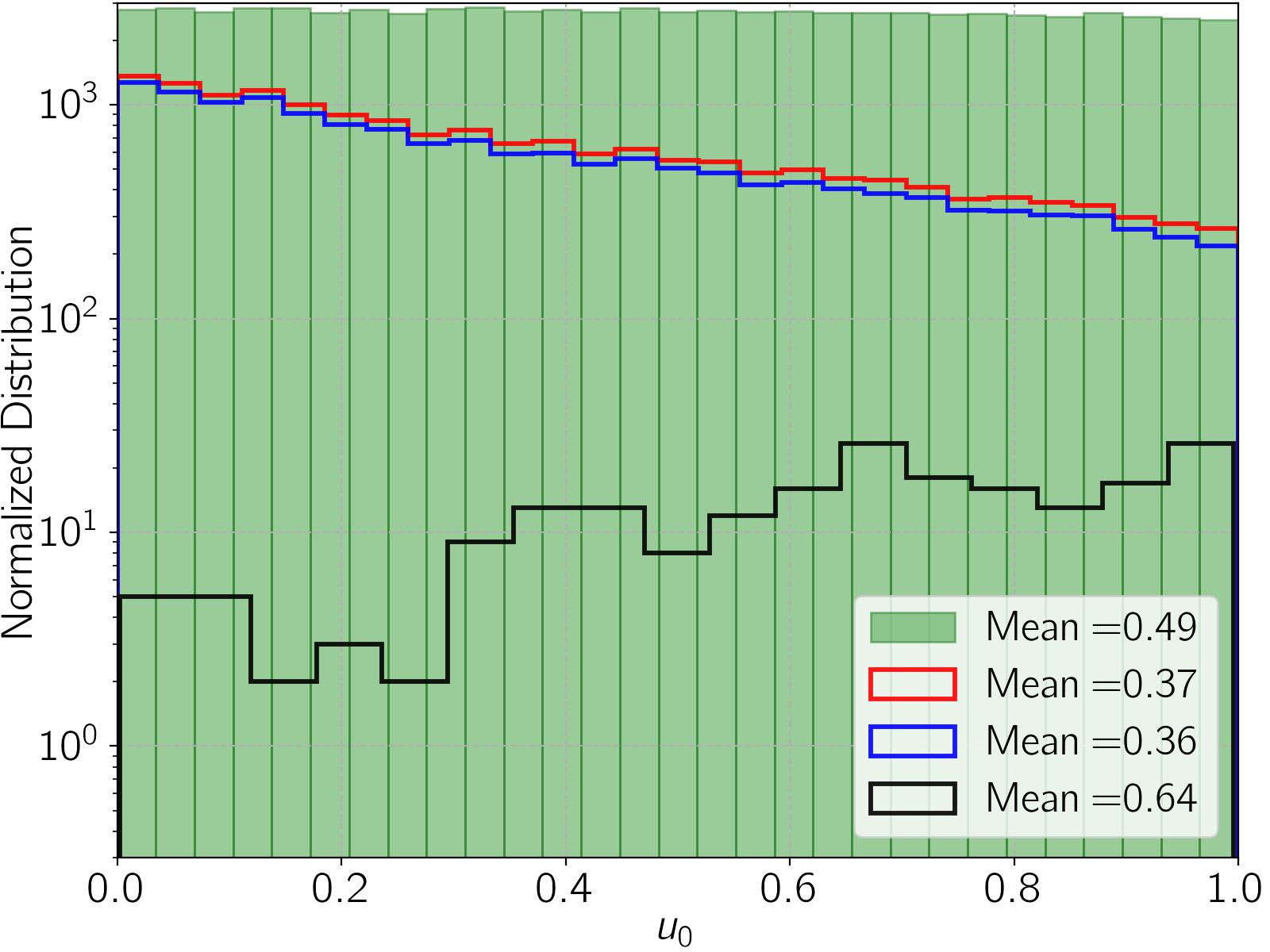}
	\includegraphics[width=0.32\textwidth]{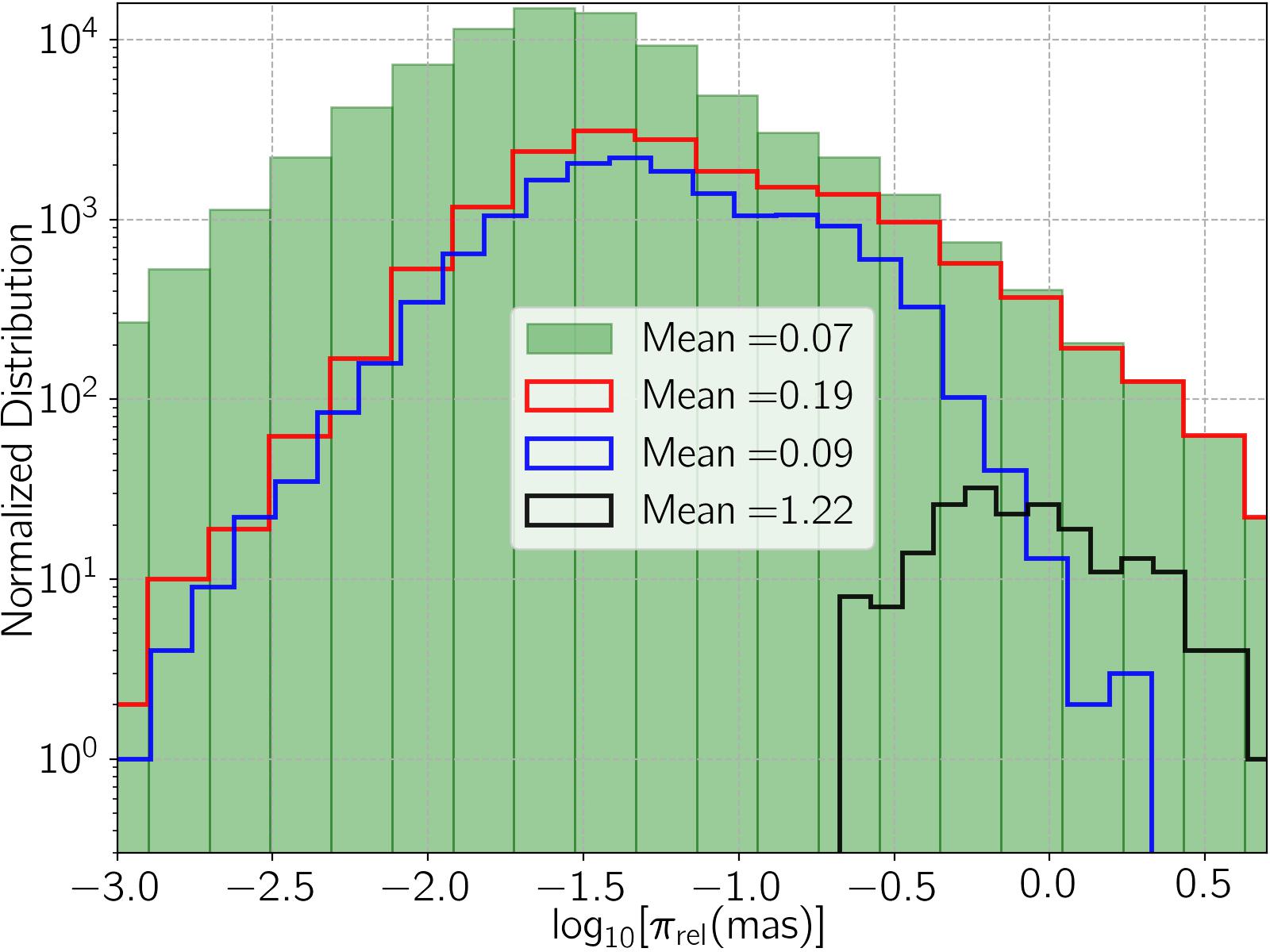}
	\includegraphics[width=0.32\textwidth]{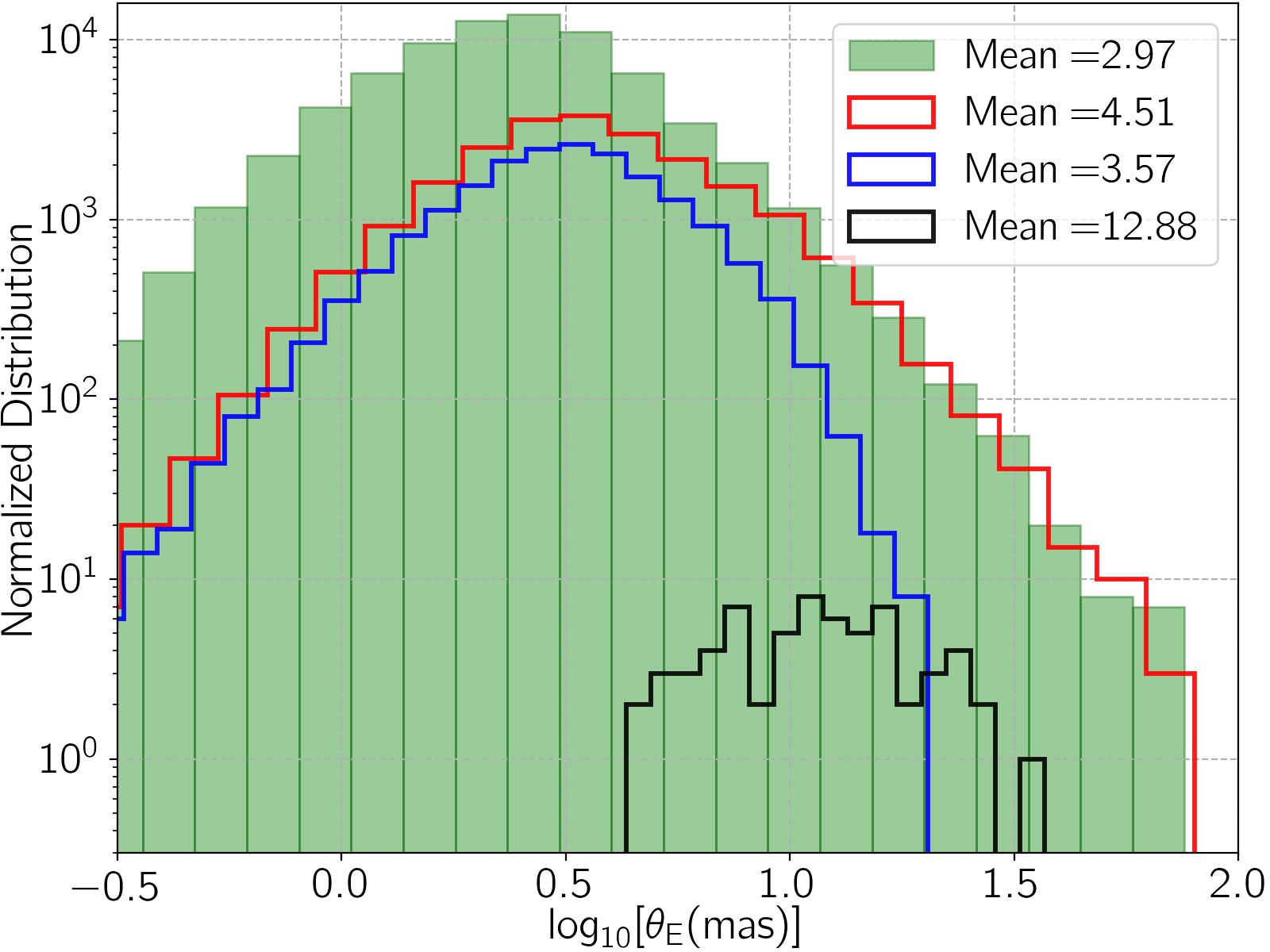}
	\includegraphics[width=0.32\textwidth]{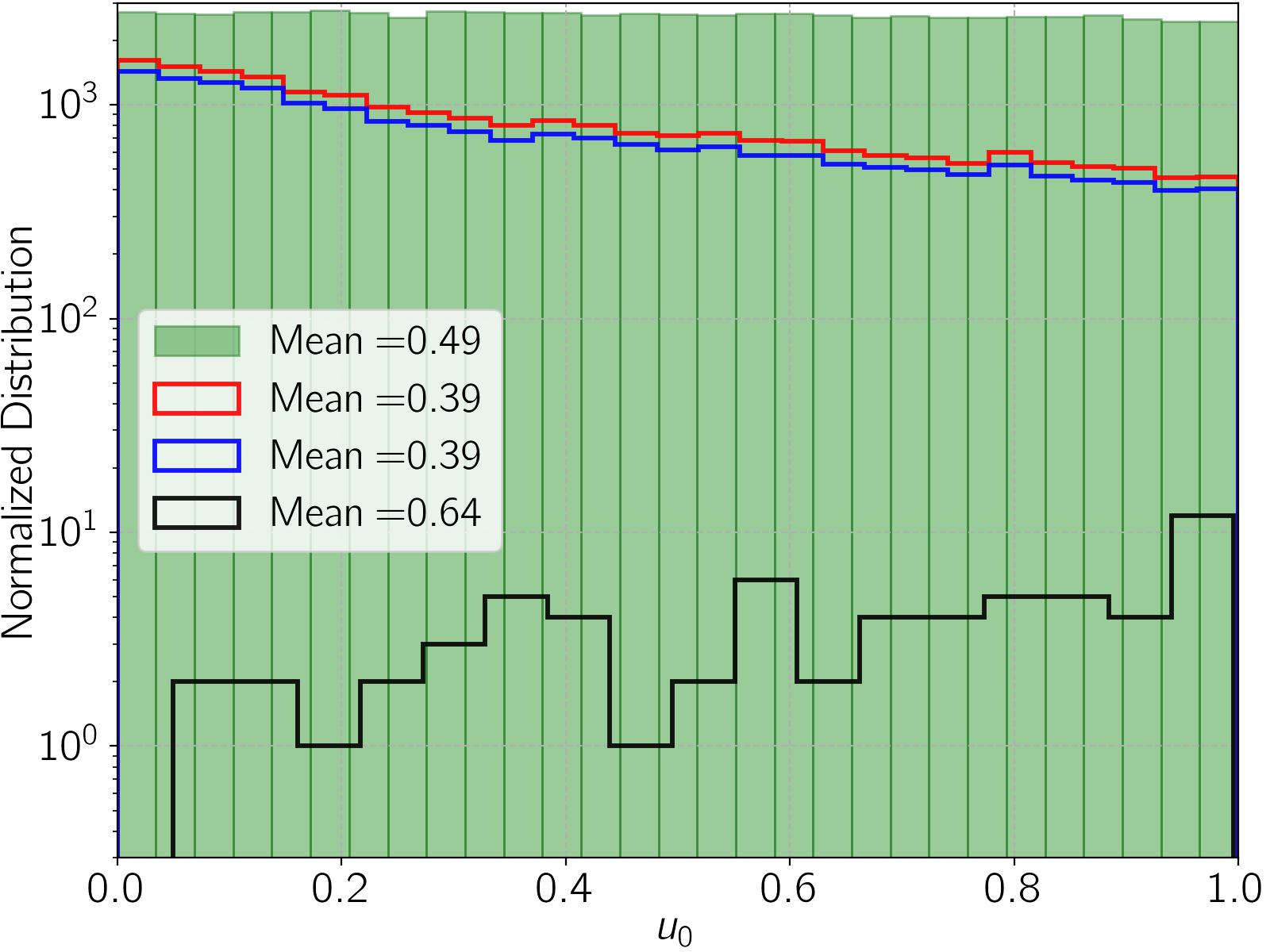}
	\includegraphics[width=0.32\textwidth]{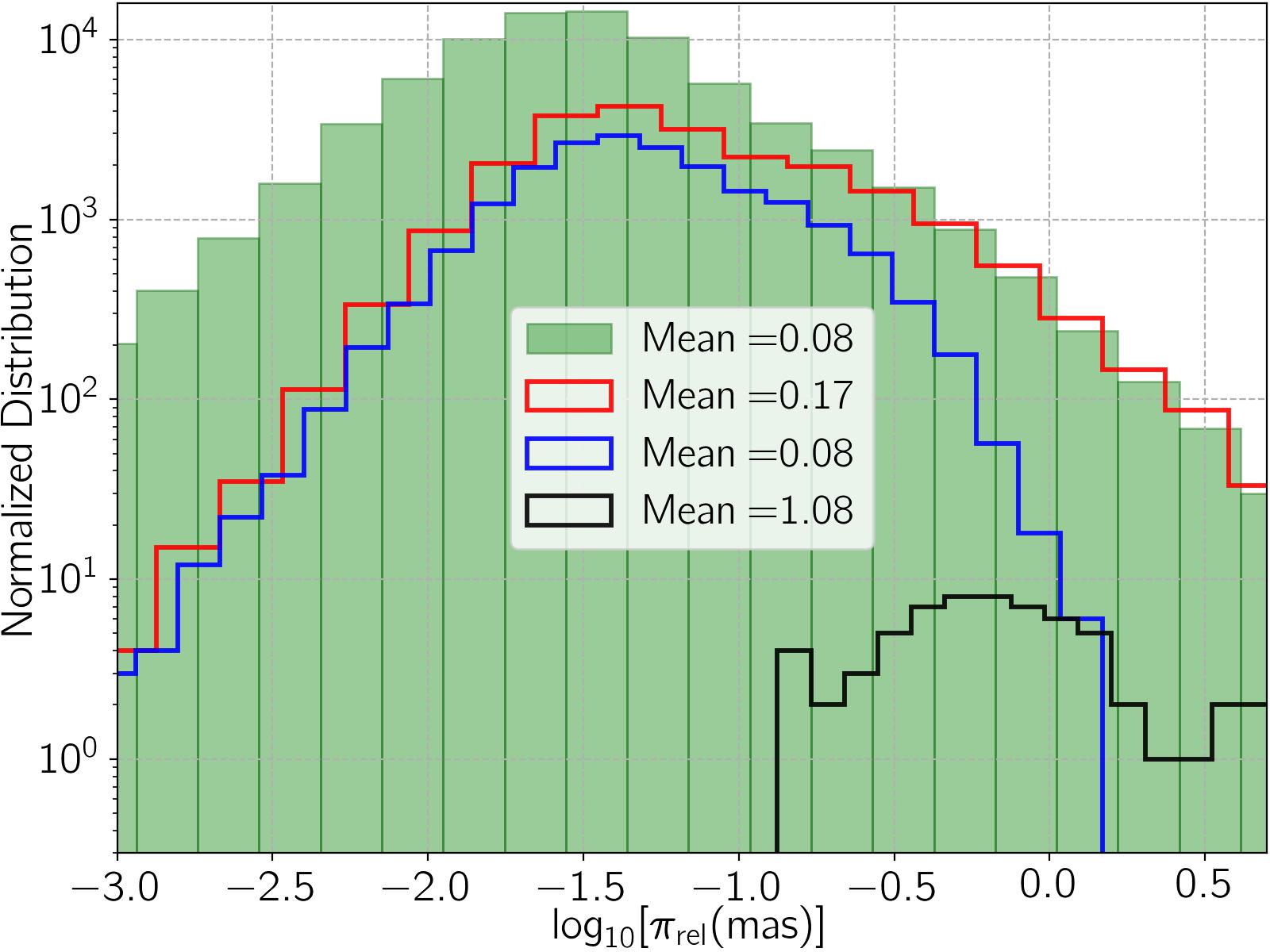}
	\includegraphics[width=0.32\textwidth]{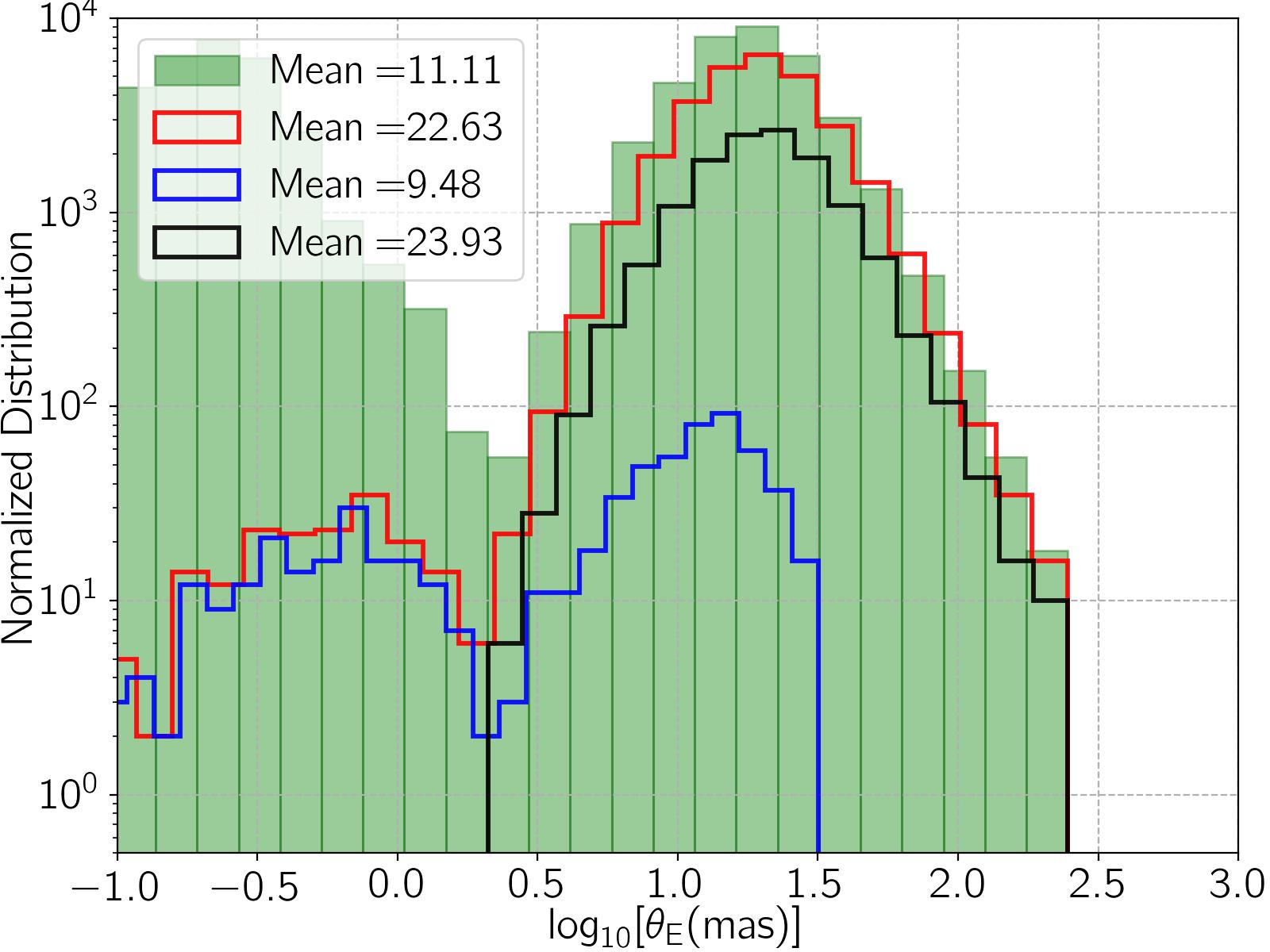}
	\includegraphics[width=0.32\textwidth]{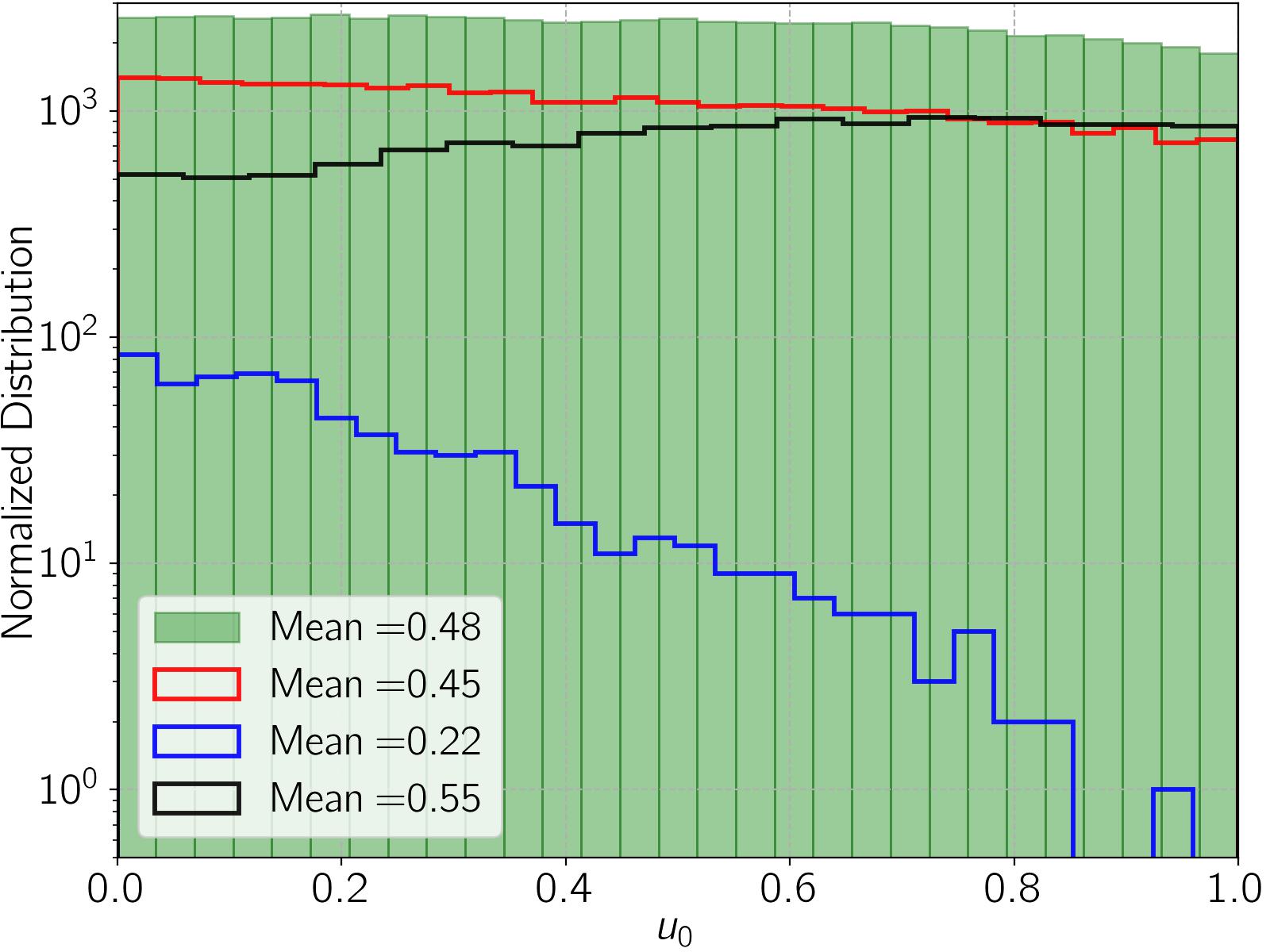}
	\includegraphics[width=0.32\textwidth]{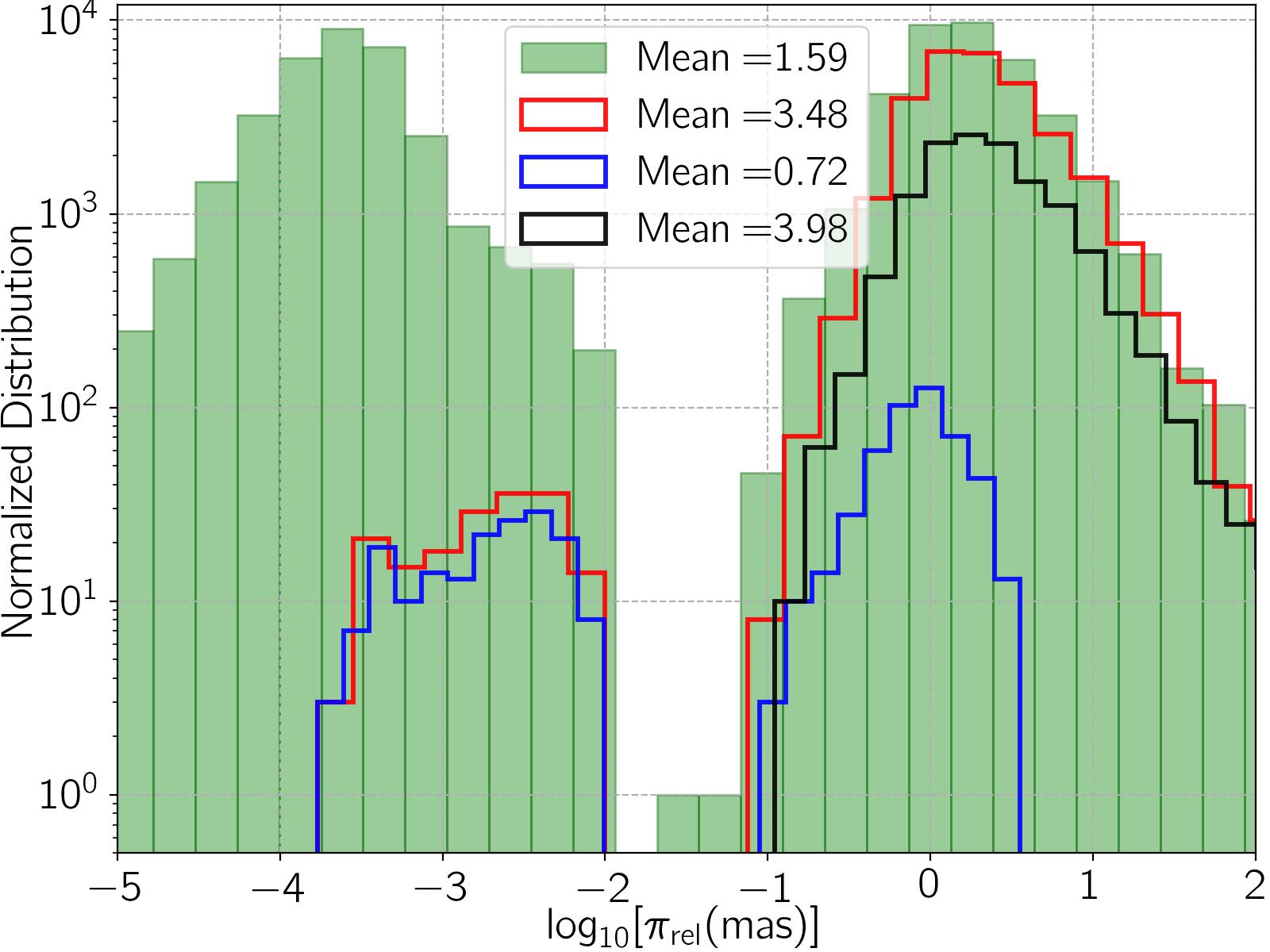}
\caption{The normalized distributions (NDs) of $\log_{10}[\theta_{\rm E}(\rm{mas})]$, $u_{0}$, and $\log_{10}[\pi_{\rm{rel}}(\rm{mas})]$ from simulations (A), (B), and (C) (from top to bottom, respectively) are represented by green color. In these panels, red step lines represent the NDs of these parameters due to events in which $\pi_{\rm E}$s can be measured (with relative error less than $4\%$) through either photometric or astrometric data. Blue and black step curves are NDs due to events with measurable parallax amplitudes through photometric data, and astrometric data, respectively. Inside each panel, the average values of parameters ($\theta_{\rm E}$,  $u_{0}$, $\pi_{\rm{rel}}$) from their distributions are indicated.}\label{NDtot}
\end{figure*}

To determine in what kind of simulated events, parallax amplitudes could be measured through astrometric observations, in three top panels of Figure \ref{NDtot} we show the normalized distributions (NDs) of $\log_{10}[\theta_{\rm E}(\rm{mas})]$, $u_{0}$, and $\log_{10}[\pi_{\rm{rel}}(\rm{mas})]$ due to events which are detectable by \wfirst\ with green filled distributions. In these figures, red step lines represent the NDs of these parameters due to the events in which $\pi_{\rm E}$s can be measured (with relative errors less than $4\%$) through either photometric or astrometric data. Blue and black step curves are NDs due to events with measurable parallax amplitudes through photometric data, and astrometric data, respectively. Inside each plot the average values of parameters from their distributions are mentioned.  

Accordingly, in the events with $\pi_{\rm E}\gtrsim 0.25 $mas (or the lens distance $D_{\rm l}\lesssim 2.7$ kpc from the observer) measuring parallax amplitudes through astrometric observations is possible. The events with larger $u_{0}$ are more favorable to infer parallax amplitudes from astrometric data, whereas in high-magnification events parallax amplitudes can be measured rather from light curves. In fact, the magnification factor tends to zero fast as $u^{-4}$, whereas the astrometric deflection maximizes when $u=\sqrt{2}$.      

{\bf B: GB Survey observations by \textit{Roman} with the ELT follow-up:~}The astrometric precision of the ELT telescope is planned to be better than 50 $\mu$as. In this simulation, we consider potential follow-up astrometric observations with the ELT telescope from long-duration microlensing events that will be alarmed by \wfirst. 

We assume this telescope will take one data every ten days, which is suitable for long-duration microlensing events. For each microlensing event the ELT observation is started when its magnification factor reaches $1.34$. 

For the ELT astrometric observations, we assume it observes microlensing events in $K$-band. In the regard of the ELT astrometric precision, the statistical astrometric accuracy of a telescope with the aperture $D$, and in the observing wavelength $\lambda$ is given by \citep[see, e.g., ][]{2010trippe}: 
\begin{eqnarray}
\sigma_{\rm a,~1}= \frac{\lambda}{\pi~ \rm{D}}\frac{1}{\rm{SNR}}= 34 \mu\rm{as} \frac{\lambda}{2.2 \mu m}\frac{42 m}{\rm{D}}\frac{100}{\rm{SNR}},
\end{eqnarray}
where, SNR is the signal to noise ratio. Hence, for the ELT observations with SNR$=100$ in $K$-band by the Multi-adaptive optics Imaging CAmera for Deep Observations (MICADO\footnote{\url{https://elt.eso.org/instrument/MICADO/}}, \citet{micado2021}) camera, the statistical astrometric precision will be $\simeq 34\mu$as. We add the systematic error ($\sim 10 \mu$as) to this statistical accuracy. Therefore, for a source star with the apparent magnitude $m_{K}\simeq 18$-$19$ mag, the MICADO astrometric accuracy reaches $\sim50~\mu$as. \citet{2020ELTaccuracy} have modeled the images would be taken by MICADO/ELT from globular clusters and found a more practical relation between the apparent magnitude of stars in the $K$-band filter and the MICADO astrometric precision adjusted for a $20$-min exposure time, which was given in their second table. Throughout the paper, we use their results to estimate the ELT astrometric accuracy in the $K$-band filter.  

We consider a $5$-month seasonal gap annually for ELT. Also, this telescope can take data from stars with the $K$-band apparent magnitude in the range $m_{K} \in [12,~23]$ mag (its saturation and detection thresholds). We assume both probabilities for suitable weather, and doing regular observations with the ELT telescope are $90\%$.   

In this simulation, extracting lensing parameters and simulating the \wfirst\ data are similar to ones explained in Simulation (A).

\begin{figure*}
\centering
\includegraphics[width=0.49\textwidth]{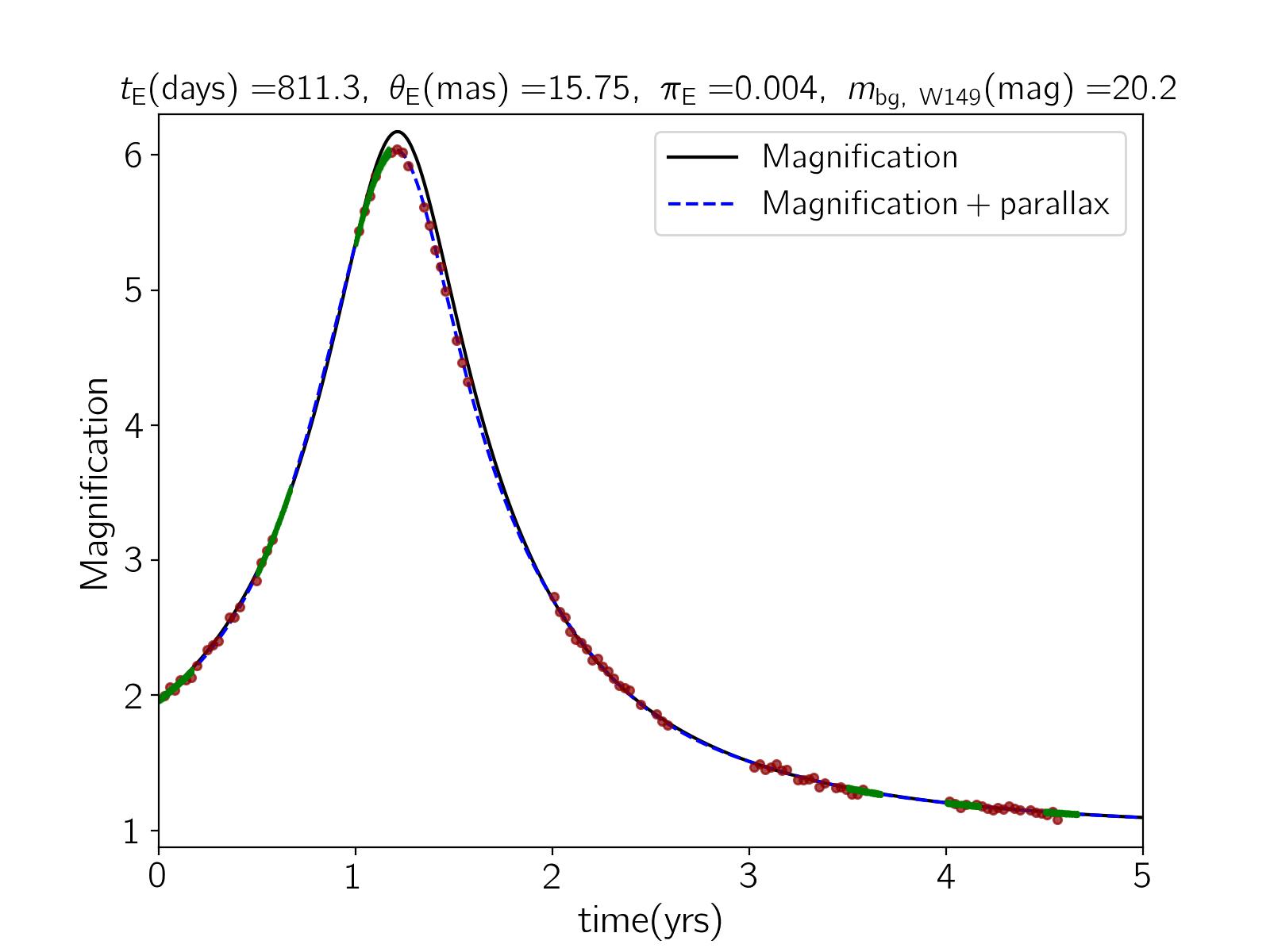}
\includegraphics[width=0.49\textwidth]{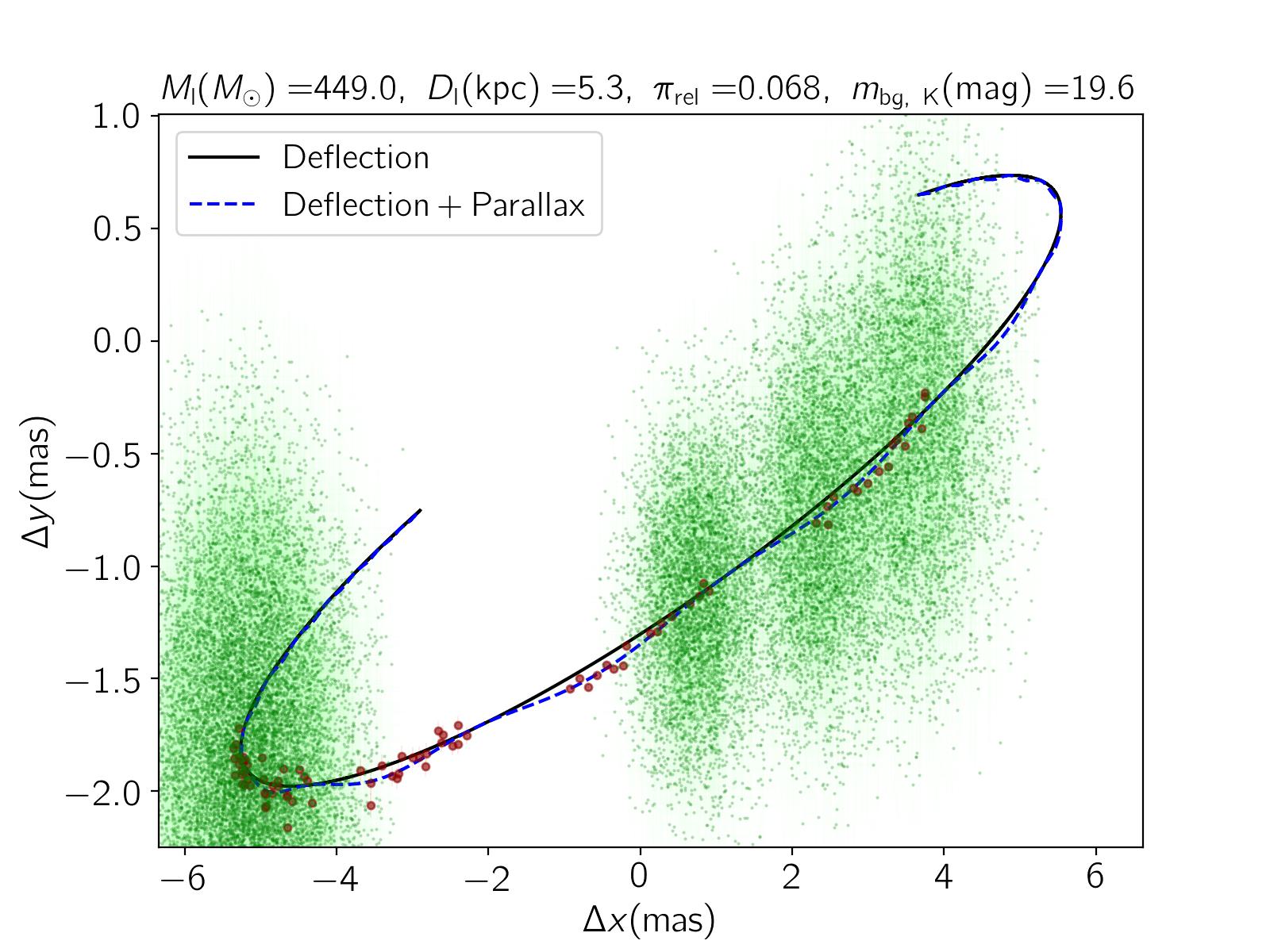}
\includegraphics[width=0.49\textwidth]{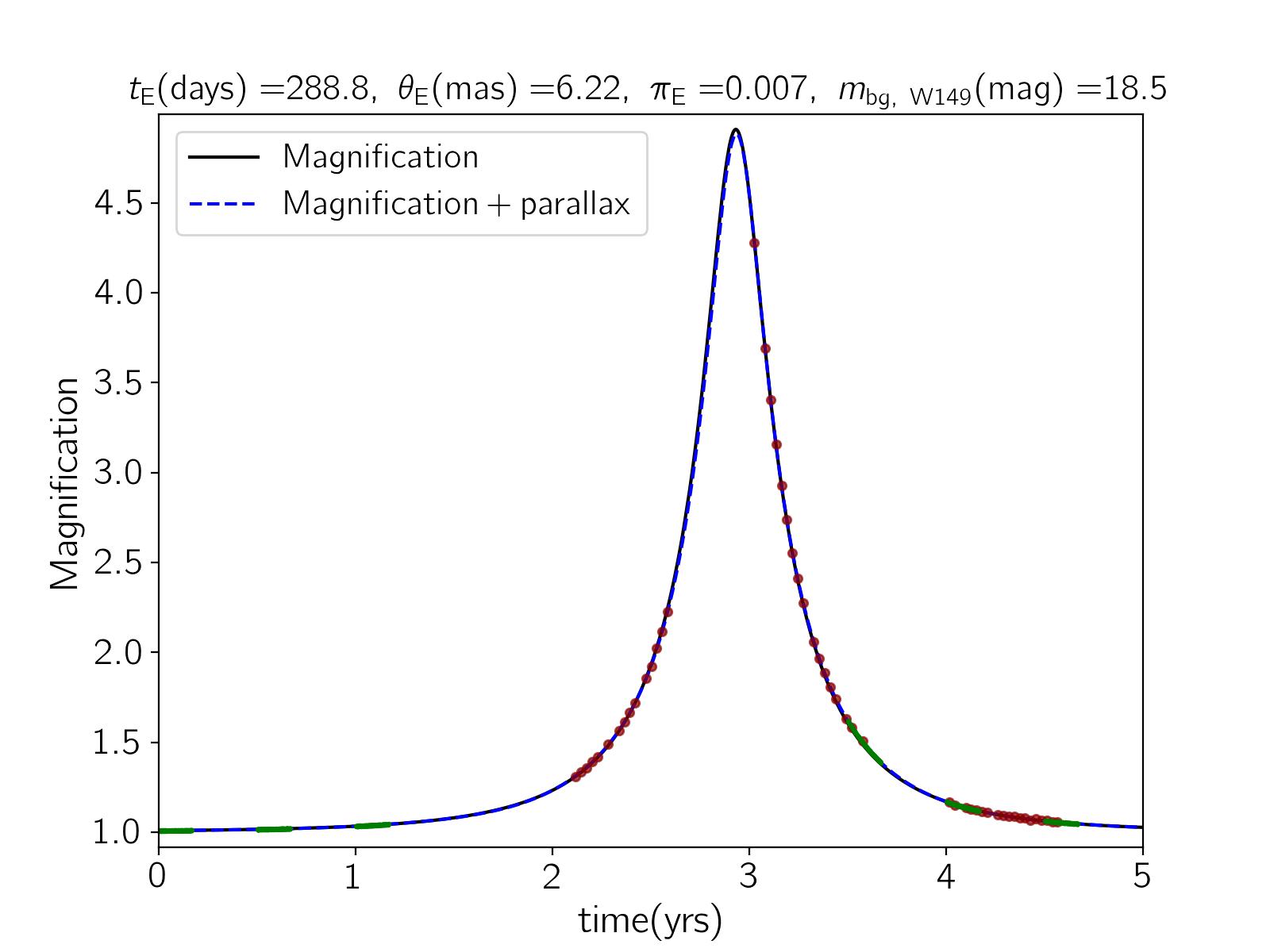}
\includegraphics[width=0.49\textwidth]{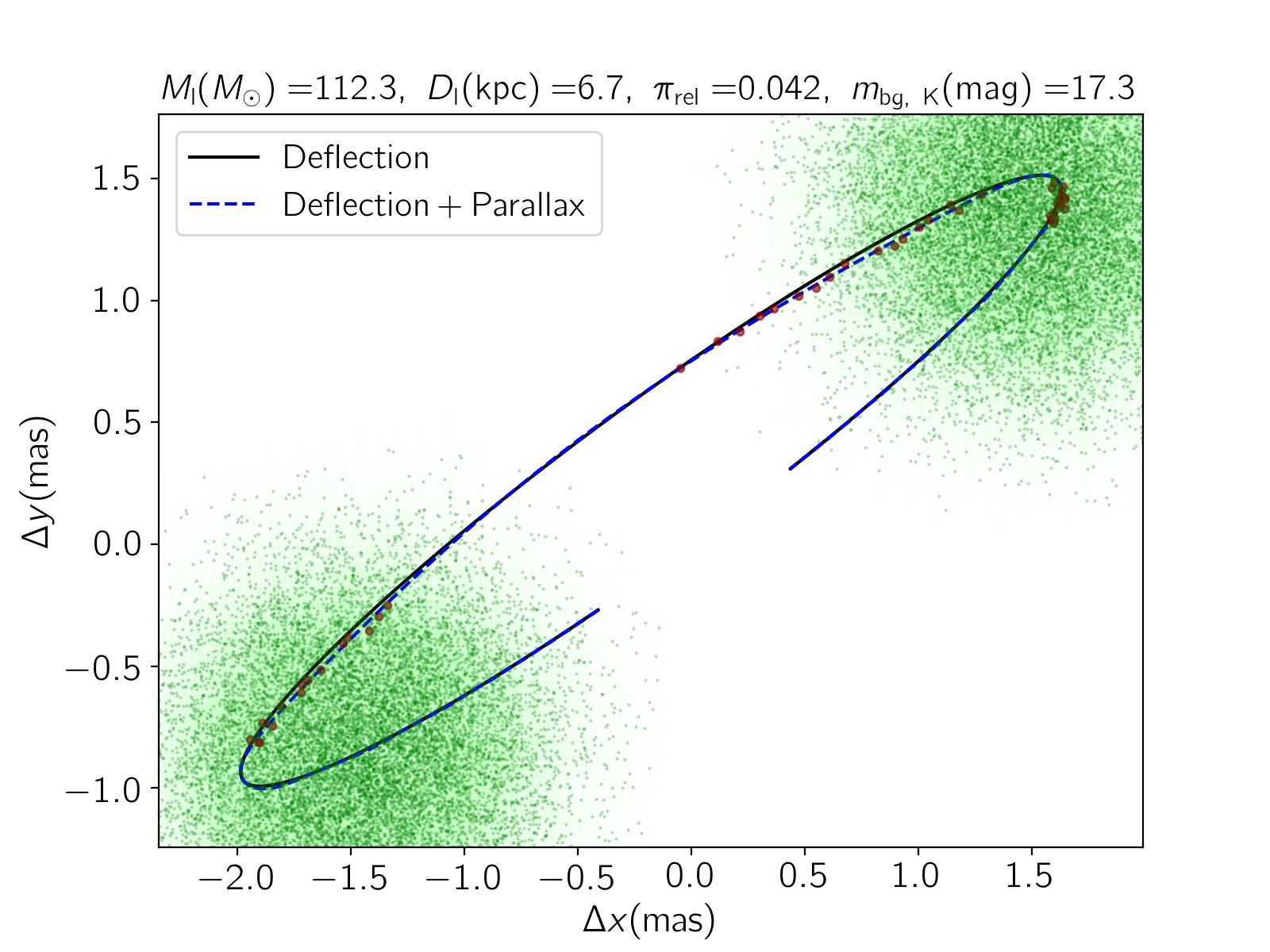}
\caption{Same as Figure \ref{Roman1}, but dark red points are hypothetically taken by the ELT telescope in the $K$-band with a $10$-day observing cadence. }\label{Roman2}
\end{figure*}

In Figure \ref{Roman2} we show two examples of simulated astrometric microlensing events detected with \wfirst\ and ELT. Here, dark red points are taken with the ELT telescope. Comparing Figures \ref{Roman1} and \ref{Roman2}, one can find the large difference between astrometric accuracies of \wfirst\ and ELT. Although the ELT data are sparse, they manifest astrometric deflections well. In the first event, discerning the parallax effect in the astrometric deflection is doable because of the ELT astrometric data.  

We mention the results from performing Monte Carlo Simulation (B) in Table \ref{tab2}. Accordingly, the ELT observations will improve the efficiency in measuring both $\theta_{\rm E}$, and $\pi_{\rm E}$. Although the efficiency for discerning parallax amplitudes from astrometric deflections is low for observations toward the Galactic bulge, some sparse data points with the ELT telescope double this efficiency. This improvement can be noticed by comparing the two first panels of Figure \ref{sigmapi}.

In the middle panels of Figure \ref{NDtot}, the normalized distributions of $\log_{10}[\theta_{\rm E}\rm{(mas)}]$, $u_{0}$, and $\log_{10}[\pi_{\rm{rel}}(\rm{mas})]$ resulted from Simulation (B) are depicted.  They are similar to top panels (owing to Simulation (A)) with small changes.

\begin{figure*}
\centering
\includegraphics[width=0.49\textwidth]{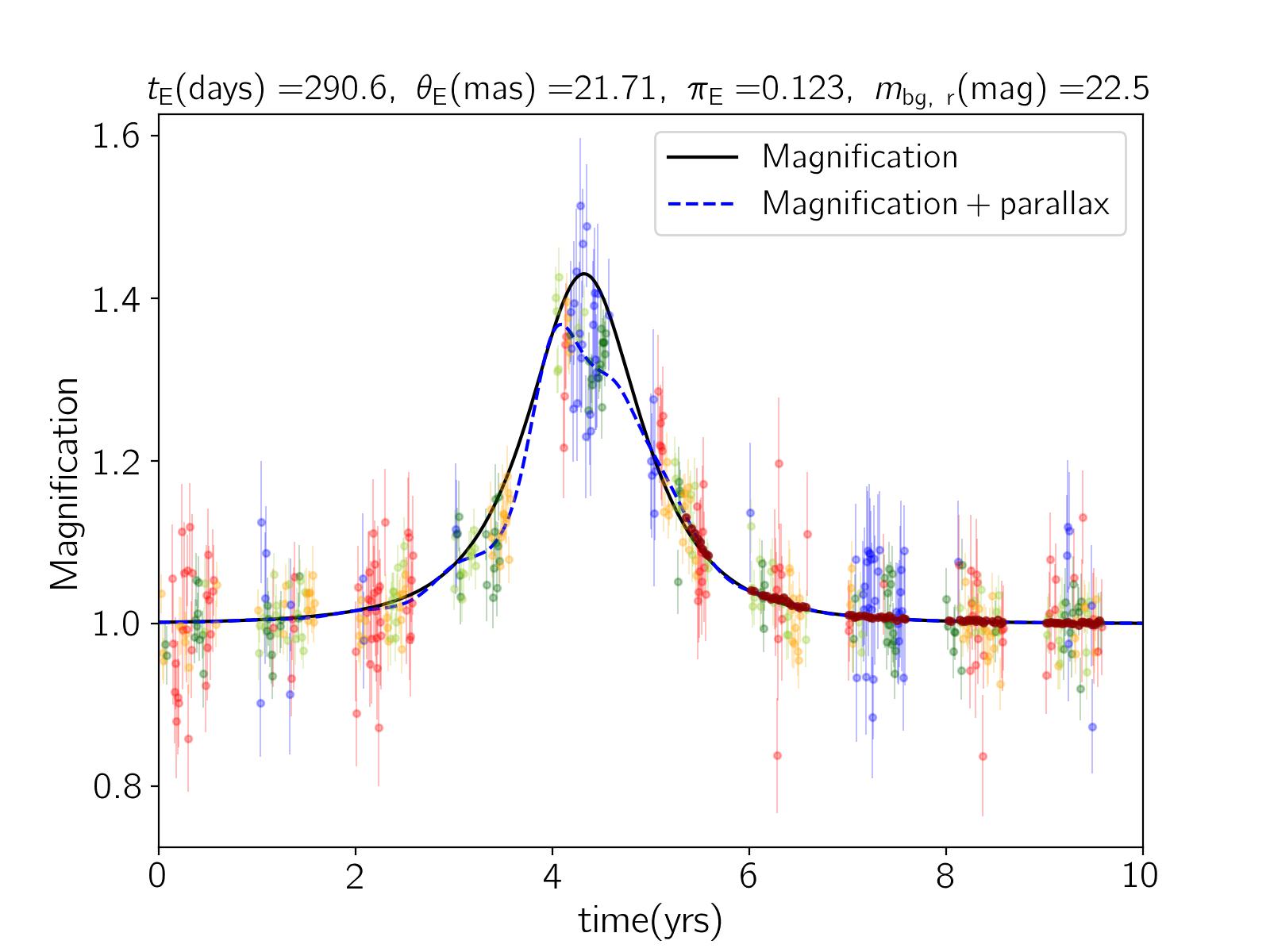}
\includegraphics[width=0.49\textwidth]{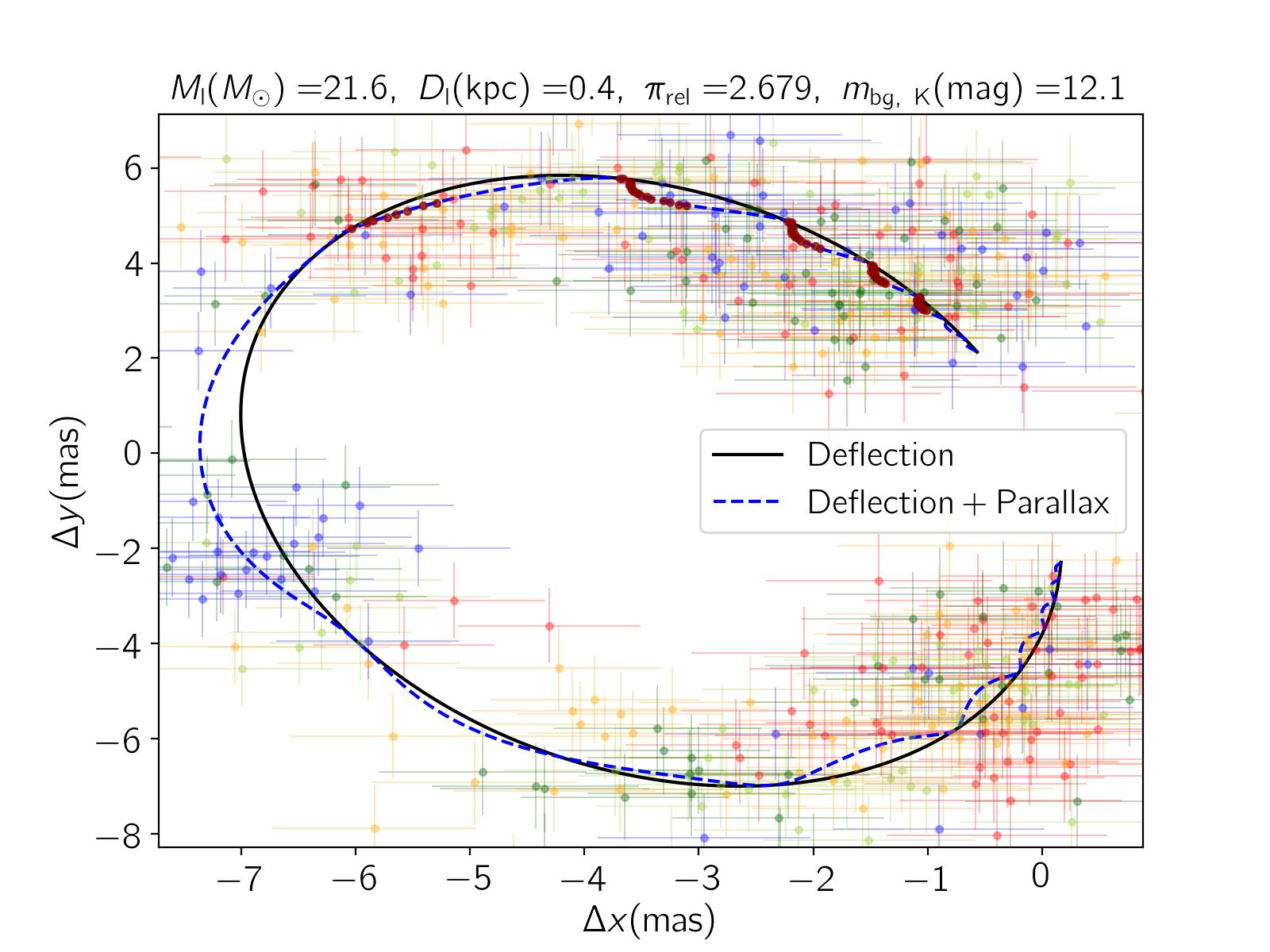}
\includegraphics[width=0.49\textwidth]{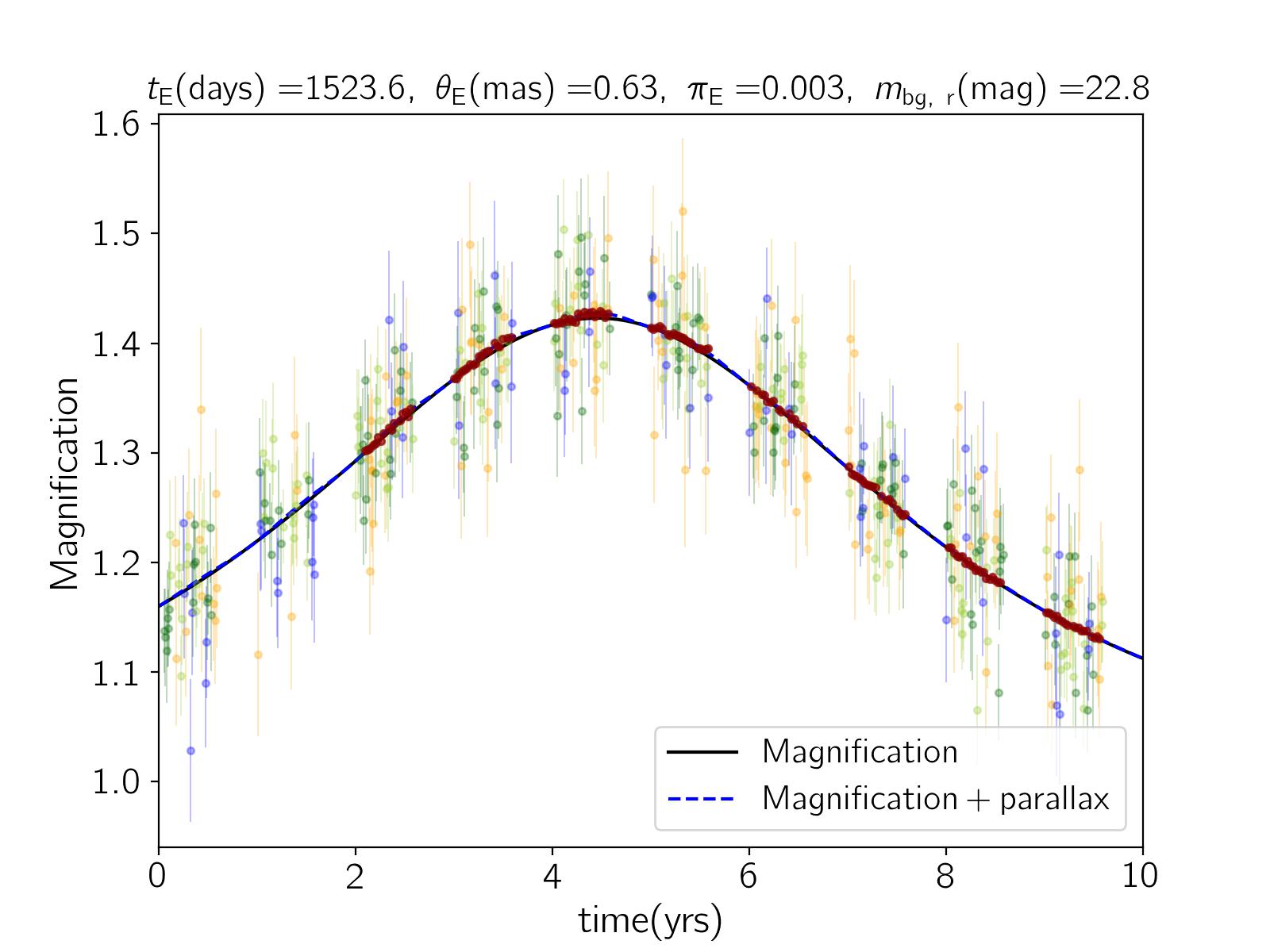}
\includegraphics[width=0.49\textwidth]{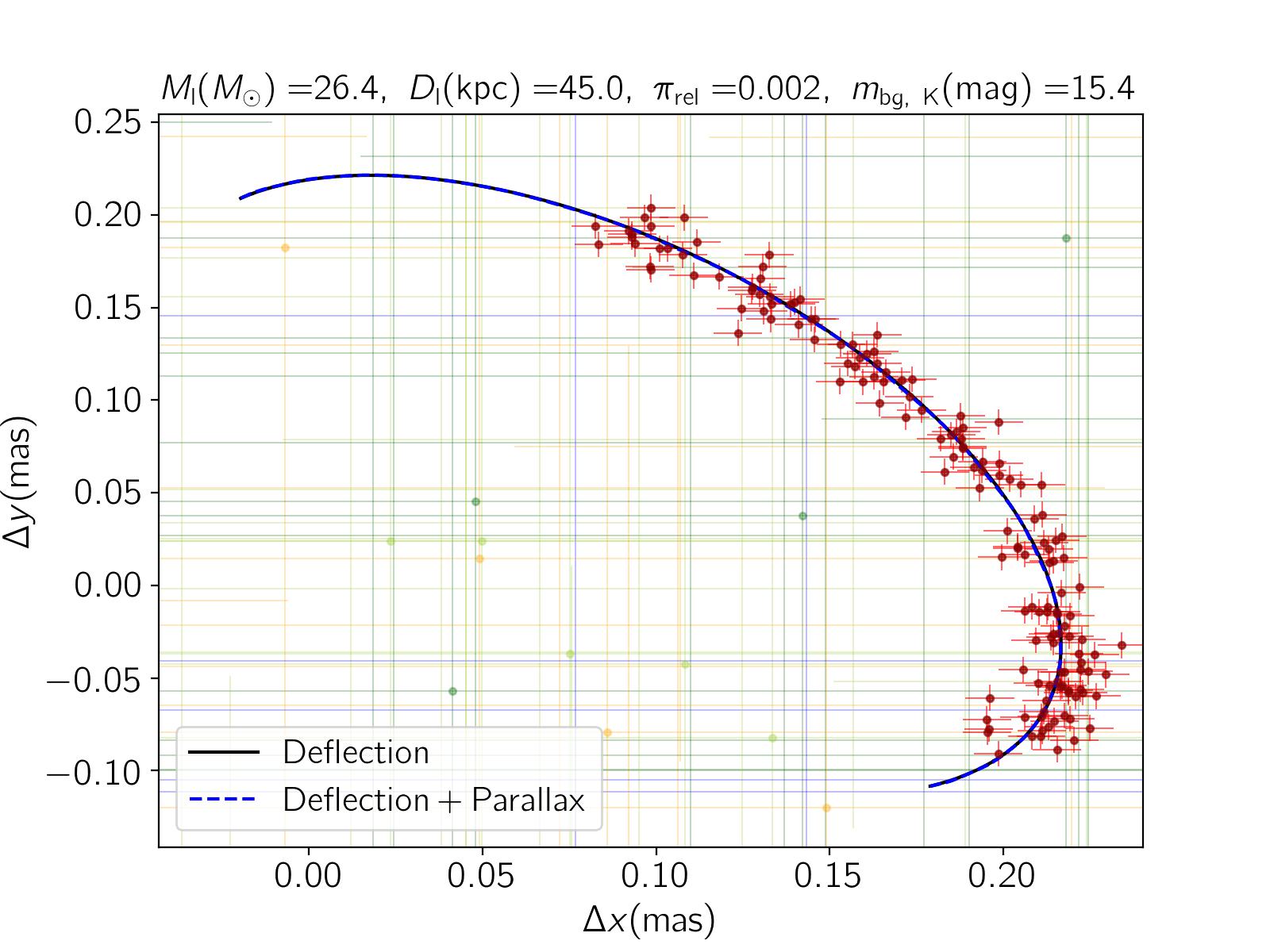}
\caption{Same as Figure \ref{Roman1}, but for observations toward the LMC with LSST. The simulated data points taken by LSST in $ugrizy$ filters are shown with purple, blue, dark green, yellowish green, orange, and red, respectively. The data taken with ELT are depicted with dark red color. The observing cadences for the LSST and ELT data are 3 and 10 days, respectively.}\label{LSST}
\end{figure*}
{\bf C: Survey and follow-up observations with LSST and ELT: }
The upcoming LSST telescope which is under construction in the Vera C. Rubin observatory, Chile is planned to survey the whole sky with a $3$-day cadence \citep{lsstbook}. This telescope will observe LMC during its $10$-year mission and potentially detect several long-duration microlensing events in this direction. 

For the LSST observations toward the LMC, we assume follow-up astrometric observations with the ELT telescope with a $10$-day cadence. The details of this follow-up astrometric observation are the same as those explained in Simulation (B).    

The LSST data will be taken in 6 filters $ugrizy$ which are similar to the filters used by the Sloan Digital Sky Survey (SDSS) telescope \citep{Fukugita1996,lsstbook}. We determine the photometric errors in these filters using the relations explained in Section (3) of \citet{lsstbook}. The astrometric accuracy of LSST is also a function of stellar apparent magnitude \citep[see, e.g., ][]{2012Eyer}. We assume that both probabilities of suitable weather for observations and doing regular observations with LSST are $90\%$. The LSST seasonal gap lasts $5$ months.   

Since, the LSST observing time is $10$ years, so we ignore microlensing events with $t_{\rm E}>4000$ days. In these events magnification factors do not reach the baseline during the LSST observing time. Also, the time of the closest approach is uniformly chosen from the range $t_{0} \in [0,~10]$ years. Other lensing parameters are determined in the same way as explained in Simulation (A). We note that for the LSST observations the blending parameter is considerable, because of the LSST's observing depth \citep[see, e.g., ][]{sajadian_LSST}.   \\

In Figure \ref{LSST}, we depict two examples of simulated astrometric microlensing events detected with the LSST and ELT telescopes. The data points taken by LSST in the filters $ugrizy$ are shown with purple, blue, dark green, yellowish green, orange, and red, respectively. The data taken by ELT are shown with dark red color. 

\noindent The first event is halo-lensing and the second one is a self-lensing event. In the first one, the parallax effect makes considerable deviations in both magnification factor and the astrometric deflection. Since the ELT data are taken on the domain of light curve (because of the ELT saturation limit), they only cover the parallax-induced deviations in the astrometric deflection. In this event, the parallax amplitude can be only extracted from its astrometric deflection. In the second event (a self-lensing one) the ELT data manifest the astrometric deflection itself, and as a result, $\theta_{\rm E}$. In this event, the parallax amplitude is not measurable.

We perform this Monte Carlo simulation, denoted (C), and make a large sample of these astrometric microlensing events which are discernible in the LSST observations. The detectability criteria are the same as ones mentioned for Simulation (A). For each simulated event we numerically calculate Fisher and Covariance matrices. The results can be found in Table \ref{tab2}.

Comparing the results reported in Table \ref{tab2}, toward  the LMC  the efficiency for measuring $\theta_{\rm E}$ even by considering the ELT follow-up observations is less than that toward the Galactic bulge by $\sim 10\%$. There are three reasons. (i) For self-lensing events, $\theta_{\rm E}$ is too small (see, Figure \ref{Fig_scatter}). (ii) The LSST telescope can discern faint source stars with the $r$-band apparent magnitude $m_{r} \in [16,~24.3]$ mag, so most of source stars in detectable microlensing events (by the LSST observations) are faint. That causes the efficiencies for measuring parameters from photometric and astrometric data (e.g., $u_{0}$, $t_{\rm E}$, $\pi_{\rm E}$, $\theta_{\rm E}$) are lower than those for measuring parameters from the \wfirst\ data. (iii) The maximum number of the LSST data during its mission with a $3$-day cadence is $708$, whereas the maximum number of the \wfirst\ data for a microlensing event during six $62$-day seasons is $35,700$. While calculating Fisher and Covariance matrices, higher numbers of data points offer lower errors.

\noindent Nevertheless, efficiency for measuring $\pi_{\rm E}$ in microlensing events toward the LMC is more than that efficiency in events toward the Galactic bulge by $\sim 13\%$. Toward the LMC, extracting parallax amplitudes from astrometric deflections is more efficient than taking out them from light curves. According to last panels of Figure \ref{sigmapi}, and \ref{NDtot}, there are two kinds of events toward the LMC with very different $\pi_{\rm{rel}}$ values, i.e., self-lensing and halo-lensing ones. The second class has large $\pi_{\rm{rel}}$ values and is suitable to determine parallax amplitude from astrometric data.

Hence, through observations toward the LMC by LSST, efficiencies for extracting $\theta_{\rm E}$ and $t_{\rm E}$ is lower, and for extracting $\pi_{\rm E}$ is higher in comparison with observations toward the GB with \wfirst.\ For that reason, efficiencies of measuring these three parameters simultaneously through both observations are the same.

Three last panels of Figure \ref{NDtot} manifest that by decreasing the lens impact parameter, the efficiency to specify parallax amplitudes from light curves improves by two orders of magnitude.

\section{Conclusions}\label{result}
In microlensing events due to massive lens objects (e.g., ISMBHs) measuring parallax amplitudes from their magnification curves is a challenge. Because the parallax amplitude in these light curves decreases by the lens mass as $\pi_{\rm E}\propto 1/\sqrt{M_{\rm l}}$. Although these massive lenses make long-duration events in comparison to the Earth orbital period, they have small parallax amplitudes.  

In this work, we studied possibility of measuring parallax amplitudes from astrometric deflections instead of light curves. We found that parallax-induced deviations in astrometric deflections are proportional to $\pi_{\rm{rel}}$, but not $\pi_{\rm E}$. However, for detecting these second-order effects astrometric deflections themselves should be measurable. Hence, the events with large $\theta_{\rm E}$, and $\pi_{\rm{rel}}$ are suitable for measuring parallax amplitudes from astrometric data. The best events are long-duration microlensing events toward the LMC due to ISMBHs inside the Galactic halo. In these events on average parallax-induced deviations reach $\sim 10-100$ mas (see Figure \ref{Fig_scatter}). The largest parallax-induced deviations on astrometric deflections occur when $u\simeq \sqrt{2}$ (at light curves' domain).    

To quantitatively study detectability of parallax amplitudes in astrometric measurements, we have done three realistic Monte Carlo simulations based on upcoming microlensing surveys, as follows. (A) Galactic bulge survey microlensing observations with \wfirst\ in its $5$-year mission. In this simulation, we assumed that the \wfirst\ telescope itself would take some sparse data points (one hour observation every $10$ days) during its large seasonal gap and when the Galactic bulge is observable. (B) Galactic bulge survey microlensing observations with the \wfirst\ telescope and follow-up observations with the ELT telescope in $K$-band. We assumed that ELT would take one data point every 10 days. (C) The LSST survey observations towards the LMC during its $10$-year mission, and follow-up observations with ELT by taking one data point every $10$ days.  

\noindent In simulations, for each event we numerically calculated Fisher and Covariance matrices based on synthetic photometric and astrometric data points, and then estimated errors. We included the parallax effect in both magnification factors and astrometric deflections in source trajectories. The results from these simulations were reported in Table \ref{tab2}. 

\noindent Accordingly, for observations toward the Galactic bulge and in events due to ISMBHs at the distances $D_{\rm l}\lesssim 2.7$ kpc from the observer, measuring parallax amplitudes through astrometric deflections is possible. However, for microlensing events toward the Galactic bulge, the efficiency to measure parallax amplitudes (with relative errors less than $4\%$) from astrometric deflections is only $2\%$. This efficiency gets double by adding the ELT sparse data points. 

\noindent In the LSST observations toward the LMC, the efficiency for measuring the parallax amplitudes increases by more than $13\%$. Toward the LMC and specially in halo-lensing events, extracting parallax amplitudes from astrometric deflections is more efficient (by $18\%$) than extracting them from light curves. 

Briefly, through observations toward the LMC by LSST, efficiencies for extracting $\theta_{\rm E}$, and $t_{\rm E}$ are lower (because LSST will mostly detect microlensing events of faint source stars, and its cadence is long), and for extracting parallax amplitudes its efficiency is higher in comparison to observations toward the Galactic bulge with \wfirst.\ Hence, the efficiencies for measuring three parameters simultaneously through both observations are in the same range. 

\small
The  source codes have been developed for this work can be found in the GitHub and Zenodo addresses: \url{https://github.com/SSajadian54/Parallax_Astrometry}, and \url{https://zenodo.org/record/8342045}\citep{sajadian_sedighe_2023_8342045}.

\bibliographystyle{aasjournal}
\bibliography{paper}{}
\end{document}